\documentclass[a4paper, 12pt, onecolumn, nofootinbib,superscriptaddress]{revtex4}%\documentclass[onecolumn,12pt]{article}
\usepackage{amsfonts, amssymb, graphicx, float, appendix, multirow, longtable}
\usepackage{amsmath}
\usepackage{color}
\usepackage{hyperref}
\usepackage{mathrsfs}
\usepackage[left=2.5cm,top=2cm,right=2.5cm,bottom=3cm]{geometry}
\usepackage{subfigure}

\usepackage{booktabs}
\usepackage{siunitx}

\usepackage{tabularx}
\usepackage{array}
\usepackage{multirow}
\usepackage{threeparttable}

\usepackage{makecell}

\newcolumntype{Y}{>{\raggedright\arraybackslash}X}

\newcommand{\sfrac}[2]{{\textstyle{#1\over#2}}}

\newcommand{\dl}{\text{d}}
\newcommand{\re}{\Re}
\newcommand{\im}{\Im}
\newcommand{\dett}{\text{det}}

\newcommand{\scalaronmass}{\mu}
\newcommand{\frequencyvar}{\omega}

\begin{document}

\title{\textbf{Scalaron Quasinormal and Quasi-Bound Modes of Schwarzschild Black Holes in the Bondi–Sachs Formulation of Metric $f(R)$ Gravity}}
\author{Bishop Mongwane}
\email{bishop.mongwane@uct.ac.za} 
\affiliation{Department of Mathematics \& Applied Mathematics,
  University of Cape Town, Rondebosch 7701, South Africa}

\author{Didam G. A. Duniya}
\email{duniyaa@biust.ac.bw}
\affiliation{Department of Physics \& Astronomy,
  Botswana International University of Science and Technology, Palapye, Botswana}

\author{Seturumane Tema}
\email{temasj@unisa.ac.za}
\affiliation{UNISA Centre for Astrophysics and Space Sciences, College of Science, Engineering and Technology,  University of South Africa
Cnr Christian de Wet Road and Pioneer Avenue, Florida Park
1709 Roodeport, South Africa }

\author{Sanele Ziqubu}
\email{zqbsan001@myuct.ac.za}
\affiliation{Department of Mathematics \& Applied Mathematics,
  University of Cape Town, Rondebosch 7701, South Africa}

\author{David F. Mota}
\email{d.f.mota@astro.uio.no}
\affiliation{Institute of Theoretical Astrophysics,
  University of Oslo, P.O. Box 1029 Blindern, 0315 Oslo, Norway}

\begin{abstract}
We study scalaron quasinormal modes and quasi-bound states of Schwarzschild black holes directly within the Bondi–Sachs formulation of metric \(f(R)\) gravity, combining frequency-domain analyses with characteristic time evolution. Starting from the linearised characteristic equations, we obtain the massive scalar master equation and show that the tensor master equation takes the same form as in general relativity. For the analytic models considered here, the linear scalar spectra depend on the \(f(R)\) theory only through the dimensionless combination \(\mu M\). We compute both quasinormal and quasi-bound spectral branches using Leaver's continued-fraction method and compare them with independent literature benchmarks. To complement the frequency domain analysis, we evolve the scalaron perturbation equations using a fourth order convergent characteristic scheme. Applying matrix-pencil stability scans to the resulting waveforms, we independently recover fundamental quasinormal frequencies consistent with the continued-fraction results within the measured extraction-stability scales. Finally, for a selected finite-mass case, we demonstrate the extraction of both oscillation frequencies and damping rates of the fundamental quasinormal mode and quasi-bound state from separate time windows of the same waveform at one extraction radius, without imposing either spectral branch at the finite outer boundary. 
\end{abstract}

\maketitle

\section{Introduction}

General Relativity (GR) remains our leading fundamental theory to describe the geometric properties of spacetime, having passed precision tests from Solar-System experiments to binary pulsars and strong-field gravitational-wave observations \cite{Will:2014kxa,PhysRevLett.116.061102}. Despite these remarkable successes, there are compelling theoretical and phenomenological motivations to consider extensions of the Einstein--Hilbert action. A broad class of modified gravity theories arises from effective field theory expansions of the Einstein–Hilbert action, incorporating higher-curvature invariants, e.g.\ \(R^2\), \(R_{\mu\nu}R^{\mu\nu}\), \(R_{\mu\nu\alpha\beta}R^{\mu\nu\alpha\beta}\), and derivative terms like \(R\Box R\). Such terms  appear generically as quantum corrections in semiclassical gravity and as $\alpha$-corrections in string-inspired effective actions, typically suppressed by a high-energy cutoff scale \cite{Stelle1978,Zwiebach:1985uq}. In the context of Cosmology, such corrections can drive early-time acceleration (e.g. Starobinsky's \(f(R)=R+\alpha R^2\) inflation) and offer late-time alternatives to dark energy \cite{STAROBINSKY198099,Sotiriou:2008rp,DeFelice:2010aj,Clifton:2011jh}. In this work we focus on metric \(f(R)\) gravity, a minimal fourth-order extension in which the Einstein-Hilbert Lagrangian $R$ is replaced by a function $f(R)$. The theory's additional scalar degree of freedom defines the sector whose perturbations we study here.

Quasinormal modes (QNMs) describe the characteristic damped oscillations of a perturbed black hole. Their complex frequencies are determined solely by the background geometry and field content, and are thus often viewed as spectral lines of the spacetime, in direct analogy with atomic spectroscopy \cite{Baibhav:2018rfk,Berti:2009kk}. With the advent of gravitational-wave astronomy, these modes have transitioned from theoretical constructs to observable quantities: the ringdown signals detected by LIGO and Virgo~ \cite{PhysRevLett.116.061102,Abbott:2016nmj} are consistent, within current signal-to-noise limitations, with the dominant QNM predictions of GR, and formally establishes the use of ringdown spectroscopy as a probe of strong-field gravity \cite{LIGOScientific:2016lio,LIGOScientific:2019fpa}. In this context, accurate theoretical predictions for QNM frequencies and damping times are essential both for interpreting observations and for studying possible signatures of modified gravity. In parallel, the spectrum of massive scalar perturbations of Schwarzschild black holes has been investigated extensively. The massive-scalar problem also admits quasi-bound states (QBSs), distinguished from outgoing QNMs by a spatially decaying condition at infinity \cite{Simone_1992,Ohashi:2004wr,PhysRevLett.109.081102}. For the class of metric \(f(R)\) models considered here, the linear scalaron obeys a massive scalar equation, with the theory dependence entering through its effective mass \(\mu\). In this work we formulate and validate the corresponding spectra directly within the Bondi–Sachs framework, providing quantitative benchmarks for characteristic evolutions.

Over the past few decades, a variety of complementary methods have been developed to compute black-hole quasinormal frequencies, which together constitute a robust toolkit for QNM computations. Semianalytic Wentzel--Kramers--Brillouin (WKB) schemes are quick and insightful in the eikonal regime and have been pushed to high orders and Pad{\'e}-resummed variants \cite{1985ApJ...291L..33S,1987PhRvD..35.3621I,Konoplya:2003ii,Matyjasek:2017psv}, while the asymptotic iteration method (AIM) is a flexible eigenvalue solver for many effective potentials \cite{2003JPhA...3611807C,Saad:2003vhv}. For high accuracy across a broad parameter space, Leaver continued-fraction methods remain a gold standard in the frequency domain, with Nollert's improvement improving convergence for highly damped overtones \cite{1985RSPSA.402..285L,1993PhRvD..47.5253N}. In this work, we employ Leaver's continued-fraction method in the frequency domain and characteristic evolution in the time domain, with the resulting waveforms analysed through matrix-pencil scans. This combination pairs explicitly imposed spectral conditions with an independent time-domain calculation, thus allowing the frequency-domain spectra to be checked against the modes recovered from the evolved waveforms.

The characteristic (Bondi--Sachs) approach foliates spacetime by outgoing null hypersurfaces and transports data to future null infinity \(\mathscr{I}^+\) \cite{Bondi21,Sachs103,lrr:winicour}. Geometrically, this is a natural setting for radiation as null hypersurfaces align with the characteristics of the field equations. Moreover, the formalism yields gauge-invariant wave-extraction at \(\mathscr{I}^+\) from numerically generated spacetimes in an unambiguous manner. Codes employing the characteristic formulation have made it possible to demonstrate the versatility of Characteristic methods in numerical relativity and have found extensive applications in, for example, binary black hole mergers \cite{Bishop:1997ik,Reisswig:2009us,Babiuc:2010ze,Handmer:2015dsa}, stellar core collapse \cite{Siebel:2003sp,Reisswig:2010cd,Ott:2010gv}, Einstein-Klein-Gordon systems \cite{Papadopoulos:1996pr,PhysRevD.54.4719,Barreto:2004fn}, observational cosmology \cite{vanderWalt:2010zd,vanderWalt:2011jt,Bester:2013fya} among others. These systems represent potential astrophysical laboratories for testing GR in the non-linear regime. The formulation has also found increasing use in an analytical setting as a tool to study linearised perturbations of black holes \cite{Bishop:2004ug,Bishop:2009ba,PhysRevD.87.104016,cedeno:2015ucy,Mongwane:2024vao}. A dedicated characteristic formulation for metric \(f(R)\) gravity was presented in \cite{mongwane:2017}. In this work, we build on this formulation to study linearised perturbations of a Schwarzschild black hole both in the frequency domain and time domain, with emphasis on the scalar mode. For the time domain treatment, we use the characteristic hierarchy on a finite radial domain and analyse the response at finite extraction radii.

Within the literature, the connection between massive-scalar spectra and the corresponding time-domain response has also been investigated directly. For example, ~\cite{PhysRevLett.109.081102} determines the quasi-bound content excited by localized scalar initial data on Schwarzschild and compares the resulting spectral superposition with numerical evolution. In addition, ~\cite{Decanini:2015yba} identifies both QNM and quasi-bound contributions in massive-scalar waveforms generated by a particle plunging into a Schwarzschild black hole. These studies provide the context for examining the recovery of both mode families within the Bondi--Sachs formulation of linear scalaron perturbations in metric $f(R)$ gravity. For a selected case, we demonstrate the recovery of the complex frequencies of the fundamental QNM and QBS from separate time windows of the same waveform at a fixed extraction radius, using a single characteristic evolution. Comparison with continued-fraction benchmarks provides a quantitative check of the recovered frequencies, obtained without imposing a spectral condition that selects either mode family at the evolution's finite outer boundary.

The scalaron master equation is obtained directly from the linearised Bondi-Sachs $f(R)$ system, and hence the subsequent frequency and time-domain analyses remain within a formalism native to the characteristic formulation. Key features of the current work include, (i) we make explicit the spectral degeneracy at fixed $\mu M$ among analytic metric $f(R)$ models on a Ricci-flat Schwarzschild background, (ii) we express the scalar QNM and QBS conditions as branch-dependent Wronskians and implement these conditions through continued-fraction equations; (iii) we provide a quantitative benchmark set against independent frequency-domain calculations, and (iv) Finally, we provide characteristic time-domain validation of both scalar spectral branches, including their recovery from a single finite-radius waveform.

This paper is structured as follows: In \S\ref{sec:metric_fr_gravity} we review Metric $f(R)$ gravity and present the field equations of the theory. In \S\ref{sec:formalism} we review the Bondi-Sachs coordinates and describe the perturbative framework employed in the derivations of the master equations. In \S\ref{sec:master_equations} we outline the derivation of the master equations for both scalar and tensor perturbations, show that the tensor equation reduces to its GR form, and summarize the scalar endpoint behaviour needed to define the spectral problems. After defining the relevant mode families and their associated boundary conditions in \S\ref{sec:mode_definitions_and_boundary_conditions}, the paper then splits the treatment in the frequency domain \S\ref{sec:frequency_domain_formulation_and_numerical_setup} and time domain \S\ref{sec:time_domain_formulations_and_numerical_setup}. Results are presented in \S\ref{sec:results} followed by concluding remarks in \S\ref{sec:concluding_remarks}. Throughout this paper, we use Geometrized units $G=c=1$ and metric signature $(-+++)$. Moreover, in our equations, we use $r$, the Schwarzschild radial coordinate and not the \textit{tortoise coordinate} that is often used in discussions of black hole perturbations.

\section{Metric $f(R)$ gravity}
\label{sec:metric_fr_gravity}

Metric $f(R)$ gravity dates back at least to Buchdahl’s work on non-linear Lagrangians \cite{1970MNRAS.150....1B} and has found renewed interest in the contexts of inflation and dark-energy cosmology \cite{STAROBINSKY198099,Sotiriou:2008rp,DeFelice:2010aj,Clifton:2011jh}. The starting point is a generalisation of the Einstein--Hilbert action 
\begin{equation}
  \label{eq:action}
  S
  = \frac{1}{16\pi} \int \! \mathrm{d}^4x\,\sqrt{-g}\,f(R) \,,
\end{equation}
where we have restricted to the vacuum case. The field equations can then be derived by varying the action \eqref{eq:action} with respect to the metric $g_{ab}$
\begin{equation}
  \label{eq:eom}
  f' R_{ab}-\frac{1}{2}fg_{ab}-\nabla_{a}\nabla_{b}f' +g_{ab}\Box f'= 0\;,
\end{equation}
where $\Box=\nabla_{c}\nabla^{c}$ is the d'Alembertian operator and we use $f=f(R)$, primes $'$ denote differentiation with respect to the Ricci scalar $R$.
The theory itself does not prescribe a specific functional form for $f(R)$. The special case $f(R)=R$ reproduces the standard Einstein-Hilbert action and therefore corresponds to GR. An important point to note, however, is that such choices are not arbitrary. Viable $f(R)$ models must reduce to GR in the appropriate limit. 

The trace of \eqref{eq:eom} yields a dynamical equation for the scalar channel
\begin{equation}
  \label{eq:trace_eq}
3\,\Box f' -2f+f'R= 0\;.
\end{equation}
When \(f''(R_0)\neq0\), this equation supplies an additional scalar
degree of freedom, in addition to the two spin-2 polarizations of the
metric.

It is often convenient to rewrite the left-hand side of \eqref{eq:eom} in terms of the Ricci scalar and its derivatives
\begin{align}
\Sigma_{ab} &= f' R_{ab}-\sfrac{1}{2}fg_{ab}-\nabla_{a}\nabla_{b}f' +g_{ab}\Box f' \\
&=  f' R_{ab}-\sfrac{1}{2}fg_{ab}-f''\nabla_{a}\nabla_{b}R - f'''\nabla_{b}R\nabla_{a}R +g_{ab}(f'''\nabla^{c}R\nabla_{c}R + f'' \Box R)\;.
\end{align}
One of the defining features of metric $f(R)$ gravity is the presence of fourth-order derivative terms of the metric, which enter via the second covariant derivatives of the Ricci scalar. This is consistent with Lovelock's theorem which states that in four dimensions the Einstein–Hilbert action (with a cosmological constant) is the only local metric action built from curvature scalars that yields second-order field equations \cite{1971JMP....12..498L,Lovelock:1972vz,Clifton:2011jh}. By contrast, Palatini $f(R)$ gravity treats the connection as an independent variable, a feature which avoids fourth-order metric equations making the theory outside the class of purely metric theories to which Lovelock's theorem directly applies \cite{Sotiriou:2008rp,DeFelice:2010aj}.

\section{Formalism}
\label{sec:formalism}

The formalism for the numerical evolution of the gravitational field equations in null cone coordinates is well known in the literature (e.g. \cite{lrr:winicour,lrr:bishop}, see also \cite{mongwane:2017} for the $f(R)$ case). In the following, we summarise the main ideas of the formalism to provide context for the variables we use in the master equations.

\subsection{Bondi-Sachs metric}
Within the Characteristic initial value problem, we employ coordinates $(u,r,x^{A})$ based on a family of outgoing null hypersurfaces emanating from an inner worldtube $\Gamma$ denoting the inner boundary of the characteristic domain. In this system, $u$ is a retarded time coordinate labelling the hypersurfaces, $r$ is a surface area coordinate and $x^{A} (A=2,3)$ \footnote{Following the standard convention, we will generally use uppercase indices for the angular directions. These will run from $2$ to $3$.} are labels for the null rays. Then the Bondi-Sachs metric takes the form
\begin{align}
  \label{eq:bs_metric}
  ds^{2} =&-\left(e^{2\beta}\left(1+\frac{W}{r} \right)-r^{2}h_{AB}U^{A}U^{B} \right)\dl u^{2}-2e^{2\beta}\dl u \dl r \nonumber \\
  & -2r^{2}h_{AB}U^{B}\dl u \dl x^{A}+r^{2}h_{AB}\dl x^{A}\dl x^{B} \;.
\end{align}
The 2-tensor $h_{AB}$, with $h^{AB}h_{BC}=\delta^{A}_{\phantom{A}C}$, satisfies the determinant condition $\dett(h_{AB}) = \dett (q_{AB})$, where $q_{AB}$ is the unit 2-sphere metric, so that $h_{AB}$ has only two degrees of freedom.

The field equations acquire a simple hierarchical structure, upon introduction of spin weighted quantities. To this end, we introduce a complex dyad $q^{A}$ on the unit sphere satisfying  $q^{A}q_{A}=0$, $q^{A}\bar{q}_{A}=2$ and $q_{A}=q_{AB}q^{B}$ with $q^{AB}q_{BC}=\delta^{A}_{C}$ and $q_{AB}=q_{(A}\bar{q}_{B)}$ \cite{1967JMP.....8.2155G,Gomez:1996ge}. We can then represent the metric functions in terms of the spin-weighted quantities
\begin{equation}
J = \sfrac{1}{2}q^{A}q^{B}h_{AB} \qquad \text{and} \qquad U = U^{A}q_{A}\;.
\end{equation}
while the angular derivatives are expressed in terms of complex differential \textit{eth} operators $\eth$ and $\bar{\eth}$.

The full non-linear system of main equations, trivial equation and supplementary conditions in the metric $f(R)$ theory are given in \cite{mongwane:2017}. In the following, we linearise the system about a Schwarzschild background.

\subsection{Background spacetime}
\label{sec:bg_spacetime}

In outgoing null coordinates, the Schwarzschild line element can be written in Eddington–Finkelstein form as
\begin{align}
\label{eq:background}
ds^{2} = - \left(1-\frac{2M}{r} \right)\dl u^{2}-2\dl u\dl r + r^{2}q_{AB}\dl x^{A}\dl x^{B}\;,
\end{align}
with $M=0$ corresponding to flat Minkowski line element. 
In terms of the Bondi–Sachs variables, the metric (\ref{eq:background}) corresponds to
\[
J=U=\beta=0\;,\qquad W=-2M\;.
\]
The Schwarzschild geometry is Ricci-flat, i.e. the Ricci tensor $\bar R_{ab}=0$ and Ricci scalar $\bar R=0$ in this background \footnote{Since \(\bar R=0\), in what follows, we use \(R\) itself to denote the
first-order Ricci-scalar perturbation.}.

Now, evaluating the vacuum metric \(f(R)\) field equations on this background shows that Schwarzschild is admitted provided
\begin{equation}
  \label{eq:schwarzschild_admissibility}
    f(0)=0 .
\end{equation}
In the remainder of this work, we restrict to models that are analytic in a neighbourhood of \(\bar R=0\) and lie on the non-degenerate branch \(f'(0)\neq0\). The corresponding positivity condition on \(f'(0)\),
together with the conditions defining the non-tachyonic scalar sector, was stated in \S\ref{sec:metric_fr_gravity}.

We note that unlike in GR, the Schwarzschild metric is not the only spherically symmetric vacuum solution that can be admitted by a general metric $f(R)$ model \cite{Capozziello:2007id,Nzioki:2009av,Sibandze:2016agp}. However, a Jebsen-Birkhoff type theorem does hold for a restricted class of theories: Suppose $f\in C^{3}$, with $f(0)=0$ and $f'(0)\neq 0$, then the only spherically symmetric vacuum solution with $R=0$ is locally Schwarzschild, a result which persist in the linear regime \cite{Nzioki:2013lca}. So, for example, for the Starobinsky model $f(R)=R+\alpha R^2$ \cite{STAROBINSKY198099}, the result applies, whereas it does not for the inverse-curvature dark-energy model $f(R)=R-\mu^4/R$ \cite{Carroll:2003wy}, which is singular at $R=0$ and therefore lies outside the class covered by the theorem.

.

\subsection{Linear perturbations}
\label{sec:linear_perturbations}

In this section, we specialise the non-linear metric \(f(R)\) equations about the Ricci-flat Schwarzschild background. We then separate the angular and temporal dependence of the metric perturbations to arrive at a coupled system of radial ODEs.

\subsubsection{Linear reduction and theory dependence}
\label{sec:fr_perturbative_reduction}

The linearised characteristic equations used below are obtained by
expanding the full nonlinear Bondi--Sachs system of
Ref.~\cite{mongwane:2017} about the Schwarzschild background. Before
carrying out this component reduction, it is useful to identify, at the
covariant level, which derivatives of \(f(R)\) can contribute at first
order.

We linearise the equations by treating deviations from this background as first-order perturbations and therefore assign
\begin{equation}
\label{key_vars}
R, J,\bar{J},U,\bar{U},w,\beta = \mathcal{O}(\epsilon)\;,
\end{equation}
where the spin-$0$ variable $w$ is related to the Bondi Sachs variable $W$ via $W=-2M+w$. The linearization then proceeds in the usual manner by discarding all terms of order $\mathcal{O}(\epsilon^{2})$ and higher, in particular any products of the first-order quantities listed in (\ref{key_vars}).

To make the dependence on the underlying \(f(R)\) model explicit, define \(f_k\) as the $k-th$ derivative of $f$ with respect to $R$, evaluated at the Schwarzschild background value $R=0$, \(f_k\equiv f^{(k)}(0)\). Then, for the analytic models considered here, in a neighbourhood
of the background value \(\bar R=0\), we have
\[
    f(R)
    =
    f_1R+\frac{1}{2}f_2R^2+O(R^3),
\]
where we have used $f(0)=0$. Now, using \(R=O(\epsilon)\), 
\[
    f(R)=f_1R+O(\epsilon^2),
    \qquad
    f'(R)=f_1+f_2R+O(\epsilon^2).
\]
Since the background Ricci scalar is constant and vanishes, one also
has
\[
    \nabla_a\nabla_b f'(R)
    =
    f_2\bar\nabla_a\bar\nabla_bR+O(\epsilon^2),
\]
where \(\bar\nabla_a\) is the covariant derivative associated with the
Schwarzschild background.
Substitution into the metric \(f(R)\) field equations Eq \ref{eq:eom}, and using the fact that
\(\bar R_{ab}=0\) in the background, gives the covariant linearised equation
\begin{equation}
f_1\,\delta G_{ab}
+
f_2
\left(
\bar g_{ab}\bar\Box
-
\bar\nabla_a\bar\nabla_b
\right)R
=0,
\label{eq:linearised_fr_covariant}
\end{equation}
where \(\delta G_{ab}\) denotes the Einstein tensor linearised about
the Schwarzschild background. Its trace therefore reduces to
\begin{equation}
    3f_2\bar\Box R-f_1R=0.
    \label{eq:linearised_trace}
\end{equation}
For \(f_2\neq0\), this becomes
\begin{equation}
    \left(\bar\Box-\mu^2\right)R=0,  \qquad
      \label{eq:ricci_flat_scalaron_mass}
    \mu^2 = \frac{f_1}{3f_2}  = \frac{f'(0)}{3f''(0)} .
\end{equation}

Together with \(f_1>0\), the finite non-tachyonic scalar sector adopted here corresponds to \(f_2>0\). If \(f_2=0\) while \(f_1\neq0\), the undivided linearised trace equation requires \(R=0\). This infinite mass limit is to be understood as the scalar-decoupling rather than as an additional finite-\(\mu\) case, and cannot be obtained by simply setting \(f_2=0\) in the finite-\(\mu\) equations.

After division of Eq.~\eqref{eq:linearised_fr_covariant} by the
nonzero factor \(f_1\), the homogeneous linearised vacuum system
depends on the underlying theory only through
\[
\frac{f_2}{f_1}=\frac{1}{3\mu^2},
\]
while all coefficients \(f_k\) with \(k\geq3\) first contribute
beyond linear order. For example, for
\[
f(R)=a_1R+a_2R^2+a_3R^3+\cdots,
\]
one has \(f_1=a_1\), \(f_2=2a_2\), and hence
\[
\mu^2=\frac{a_1}{6a_2}.
\]
The coefficients \(a_3,a_4,\ldots\) therefore do not modify the
linear scalar QNM or quasi-bound frequencies on the Ricci-flat
Schwarzschild background.

Consequently, at fixed geometric Schwarzschild mass \(M\), two models with the same value of \(f'(0)/f''(0)\) have the same value
of \(\mu M\) and therefore the same scalar QNM and quasi-bound spectra:
\[
    \frac{f_A'(0)}{f_A''(0)}
    =
    \frac{f_B'(0)}{f_B''(0)}
    \quad\Longrightarrow\quad
    2M\omega_{\rm sc}^{(\ell n)}[f_A]
    =
    2M\omega_{\rm sc}^{(\ell n)}[f_B] .
\]
The corresponding consequence for the tensor sector is established directly from the derivation of the master equations below (\S\ref{sec:master_equations}). Taken together, these results represent a degeneracy of the homogeneous linear spectral problem, not an all-order equivalence of the underlying theories \cite{Datta:2019npq,Bhattacharyya:2017tyc}. The specific degeneracy based on the values of the theory at \(\bar R=0\)
can also be lifted on backgrounds with nonzero or non-constant Ricci curvature. In this context, a change to another Ricci-flat background does not, by itself, reintroduce dependence on \(f_k\) with \(k\geq3\) at linear order.

\subsubsection{Angular and frequency decomposition}
\label{sec:harmonic_decomposition}

Because the Schwarzschild background is spherically symmetric, the linearised equations decouple for each $(\ell,m)$ mode \cite{Martel:2005ir,1986JMP....27.1238L}, and in what follows we work with a fixed $(\ell,m)$, omitting the explicit summation over multipoles. If the background was Kerr, one would need to be more careful, i.e this clean multipole decoupling is specific to spherical symmetry. For the rotating Kerr black holes, which are only axisymmetric, a decomposition in this form generically mixes different $\ell$ and full separability instead relies on spin-weighted spheroidal harmonics in the Teukolsky formalism \cite{1973ApJ...185..635T,chandrasekhar:1983}. With this in mind, it is convenient to expand all angular dependence in the metric quantities in terms of eigenfunctions of the $\eth$ and $\bar{\eth}$ operators. For a quantity $\Phi$ of spin weight $s$, in the time domain, we have 
\begin{equation}
  \Phi(u,r,\Omega) = \sum_{\ell m}\Phi_{\ell m}(u,r){}_{s}Z_{\ell m}\;.
\end{equation}
where ${}_{s}Z_{\ell m}$ are real spin-$s$ spherical harmonics (see Eq.\eqref{eq:zlms_defined} below). Due to the mode decoupling on the Schwarzschild background, we will generally fix $(\ell,m)$ and suppress the indices, and write simply
\begin{equation}
  \Phi(u,r,\Omega) = \Phi(u,r){}_{s}Z_{\ell m}\;.
\end{equation}
In the frequency domain, one can further decompose $\Phi(u,r)$, assuming harmonic time dependence as
\begin{equation}
  \label{eq:harmonic}
  \Phi(u,r) = \hat{\Phi}(r)\re\left(e^{\rho u} \right)\;.
\end{equation}
where $\hat{\Phi}$ depends only on $r$, and $\rho=-i\frequencyvar$ is a complex frequency variable. Under this ansatz, time derivatives in the field equations become algebraic factors of $\rho$, i.e.
\begin{equation}
  \label{eq:dictionary}
  \partial_u \Phi \longleftrightarrow \rho\,\hat \Phi(r) \qquad \text{and more generally} \qquad
  \partial_u^n \Phi \longleftrightarrow \rho^n \hat \Phi(r),
\end{equation}
so the time-domain and frequency-domain forms of the equations differ only by these replacements.  In this case $\frequencyvar$ is, in general, a complex frequency, whose real part $\re(\frequencyvar)$ sets the oscillation frequency of the mode, and the imaginary part sets the growth ($\im(\frequencyvar) > 0$) or decay ($\im(\frequencyvar) < 0$) rate. More explicitly, for the fields in our problem, we have \cite{mongwane:2017,Bishop:2004ug},
\begin{subequations}
\label{eq:igenfunction_decomposition}
\begin{eqnarray}
R(u,r,\Omega) &=& \hat{R}(r)\,\re\left(e^{\rho u} \right) Z_{\ell m}\;,\\
\beta(u,r,\Omega) &=& \hat{\beta}(r)\,\re\left(e^{\rho u} \right) Z_{\ell m}\;,\\
w(u,r,\Omega) &=& \hat{w}(r)\,\re\left(e^{\rho u} \right) Z_{\ell m}\;, \\
U(u,r,\Omega) &=& \hat{U}(r)\,\re\left(e^{\rho u} \right) \eth Z_{\ell m}\;, \\
J(u,r,\Omega) &=& \hat{J}(r)\,\re\left(e^{\rho u} \right) \eth^{2}Z_{\ell m}\;. 
\end{eqnarray}
\end{subequations}
An important technical point is that we use real-valued spin $s$ harmonics ${}_{s}Z_{\ell m}$ instead of the usual complex ${}_{s}Y_{\ell m}$. This serves to avoid the mixing of $+m$ and $-m$ multipoles (for a given $\ell$) that would arise from terms such ($\eth\bar{U}+\bar{\eth}U $) and ($\bar{\eth}^{2}J+\eth^{2}\bar{J}$), if ${}_{s}Y_{\ell m}$ were used instead. The ${}_{s}Z_{\ell m}$ are defined in terms of the standard spin-weighted spherical harmonics ${}_{s}Y_{\ell m}$ as \cite{1967JMP.....8.2155G}

\begin{equation}
  \label{eq:zlms_defined}
{}_{s}Z_{\ell m} = \begin{cases}
\dfrac{i}{\sqrt{2}}\left[(-1)^{m}{}_{s}Y_{\ell m} +{}_{s}Y_{\ell\; -m} \right] & \text{ for } m < 0\;,\\
\hfill {}_{s}Y_{\ell m} \hfill & \text{ for } m=0\;,\\
\dfrac{1}{\sqrt{2}}\left[(-1)^{m}{}_{s}Y_{\ell m} +{}_{s}Y_{\ell\; -m} \right] & \text{ for } m > 0\;.
\end{cases}
\end{equation}

\subsubsection{Separated linearised system}
\label{sec:separated_linear_system}

Substituting the decompositions above into the nonlinear characteristic equations of Ref.~\cite{mongwane:2017}, retaining terms through \(\mathcal{O}(\epsilon)\), and suppressing the radial-function hats
gives\footnote{We correct a typographical omission of an $\eth$ operator in the scalaron terms (last two terms) of Eq (53b) in ref \cite{mongwane:2017}. Additionally, Eq 53a is now corrected. None of these typographical errors affect the results of \cite{mongwane:2017}, and in fact, the subsequent system (61a--e) has all the correct terms.} 
\begin{subequations}
  \label{eq:radial_system}
  \begin{eqnarray}
\label{eq:lbeta}
\beta_{,r} - \frac{r}{12\scalaronmass^{2}}R_{,rr}=0\;,
\end{eqnarray}
\begin{align}
  &\label{eq:lU}
  r^{3} U_{,rr}
  + 4 r^{2} U_{,r}
  - (\ell-1)(\ell+2)\, r\, J_{,r}
  + 4\,\beta
  - 2 r\, \beta_{,r}
  + \frac{2}{3 \scalaronmass^{2}} R
  - \frac{2 r}{3 \scalaronmass^{2}} R_{,r}
  = 0\;, \\[6pt]
  &\bigl[4 + 2\,\ell(\ell+1)\bigr]\,\beta
  + (\ell-1)\,\ell(\ell+1)(\ell+2)\,J
  - 4\,\ell(\ell+1)\,r\,U
  - \ell(\ell+1)\,r^{2}\,U_{,r}
  - 2\,w_{,r}
  - \frac{r^{2}}{3}\,R \nonumber\\
  &\hspace{4em}
  - \frac{2r}{3\scalaronmass^{2}}\!\left(1 - \frac{2M}{r}\right)\!R_{,r}
  + \frac{2\rho}{3\scalaronmass^{2}}\,R
  = 0\;, \label{eq:lw}\\[6pt]
  &\label{eq:lJ}
  2\rho\bigl(r J + r^{2} J_{,r}\bigr)
  - 2\,\beta
  + 2 r\,U
  + r^{2} U_{,r}
  - 2(r-M)\,J_{,r}
  - r^{2}\!\left(1 - \frac{2M}{r}\right) J_{,rr}
  - \frac{1}{3 \scalaronmass^{2}} R
  = 0\;, \\[6pt]
  &\left(1-\frac{2M}{r}\right) R_{,rr}
  - \frac{2}{r}\bigl(\rho\,R + r\,\rho\,R_{,r}\bigr)
  + \frac{2}{r}\left(1-\frac{M}{r}\right) R_{,r}
  - \frac{\ell(\ell+1)}{r^{2}}\,R
  - \scalaronmass^{2} R
  = 0\;.
  \label{eq:ltrace}
\end{align}
\end{subequations}
In the above, \eqref{eq:ltrace} is the closed scalar trace equation, whereas the remaining equations determine the metric perturbations. These equations provide the starting point for the scalar and tensor master equations derived in Sec.~\ref{sec:master_equations}.

\section{Master equations}
\label{sec:master_equations}

From the linearised system \eqref{eq:radial_system} we obtain two independent master equations: one for the scalar perturbations, encoded in the spin-$0$ variable $R$, and one for the gravitational perturbations, encoded in the spin-$2$ variable $J$. The trace equation \eqref{eq:ltrace} is a closed second-order ODE for $R$ and therefore directly provides the scalar master equation. By contrast, the master equation for $J$ is less trivial and requires eliminating the other variables.  We first carry out this elimination and establish the reduction to the GR tensor equation (\S\ref{sec:tensor_master_reduction}), before collecting the scalar master equation and the endpoint behaviours (\S\ref{sec:scalar_perturbations}) required for the spectral problems considered below

\subsection{Tensor reduction and GR form}
\label{sec:tensor_master_reduction}

Consider the three equations $\eqref{eq:lU}$, $\eqref{eq:lJ}$ and $\partial_r\eqref{eq:lJ}$, and regard $U$, $U_{,r}$ and $U_{,rr}$ as algebraic unknowns, with  $J,\beta, R$ and their $r-$derivatives treated as as known source functions. The remaining hypersurface relation \eqref{eq:lw} containing $w_{,r}$ plays no role in this elimination. Indeed, within this subsystem of equations, the $U$-dependence enters only through the combinations
\begin{equation}
2rU+r^2U_{,r},\qquad
2U+4rU_{,r}+r^2U_{,rr},\qquad
4r^2U_{,r}+r^3U_{,rr}.
\end{equation}
so that $\eqref{eq:lU}$, $\eqref{eq:lJ}$ and $\partial_r\eqref{eq:lJ}$ form a $3-$by$-3$ linear system for $(U,U_{,r}, U_{,rr})$. Solving this system by standard elimination yields explicit expressions for $U$, $U_{,r}$ and $U_{,rr}$ in terms of $J$ and its $r$-derivatives, together with $\beta$ and the scalar perturbation $R$. The spin--2 master equation then follows by enforcing the compatibility identity
\begin{equation}
\label{eq:compatibility}
\partial_r U-U_{,r}=0,
\end{equation}
which must hold since $U_{,r}$ is by definition the radial derivative of $U$.
Substituting the solved expressions for $U_0$ and $U_{0,r}$ into \eqref{eq:compatibility}, one finds that all terms involving $\beta$ and $R$ collapse into the single combination
\begin{equation}
\frac{2}{r}\beta_{,r} -\frac{1}{6\scalaronmass^2}R_{,rr}\;,
\end{equation}
which vanishes identically by the hypersurface relation \eqref{eq:lbeta}.  Consequently, \eqref{eq:compatibility} reduces to a closed equation for $J$ alone. As pointed out in \cite{Mongwane:2024vao}, rather than working directly with the resulting fourth order equation for $J$, it is convenient to work with the auxiliary variable $\mathcal{J}=r^3(rJ(r))_{,rr}$. In this case, $\mathcal{J}$ satisfies a second order ODE
\begin{equation}
  \label{eq:tensor_master_eq}
  r^2\left(r-2M \right)\mathcal{J}_{,rr}-2r\left(\rho r^2+r-5M \right)\mathcal{J}_{,r}-\left[2\rho r^2+(\ell^2+\ell-2)r+16M \right]\mathcal{J} = 0\;.
\end{equation}
This is the Schwarzschild tensor master equation in the Bondi--Sachs formulation of GR \cite{Mongwane:2024vao}. Since the horizon-regular and outgoing large-radius QNM conditions are also unchanged, its intrinsic QNM spectrum is the standard GR Schwarzschild spectrum. We therefore do not repeat the established tensor spectral analysis contained in \cite{Mongwane:2024vao}. Instead, we use the corresponding recurrence is only as a high-overtone consistency check of the continued-fraction implementation, with the Nollert termination and its numerical validation given in Appendix~\ref{app:tensor_validation}.

\subsection{Scalar master equation and endpoint behaviour}
\label{sec:scalar_perturbations}

As mentioned previously, the scalar sector is governed by the trace equation \eqref{eq:ltrace}, which provides a closed second-order radial ODE for the scalaron perturbation $R(r)$,
\begin{equation}
  \label{eq:scalar_master_eq}
  r\left(r-2M\right)R_{,rr}-2\left(\rho r^2-r +M\right)R_{,r}-\left[2\rho r+\ell(\ell+1)+\scalaronmass^2r^2\right]R=0\;.
\end{equation}
In the following, we summarise the local behaviour of solutions at the horizon and at infinity, since these determine the physically relevant boundary conditions used later to compute the scalar quasinormal and quasi-bound spectrum.
\begin{itemize}
  
\item \textit{Regular singular point at $r=2M$}
  \\If we express the solutions of the master equation (\ref{eq:scalar_master_eq}) near the singularity at ${r=2M}$, as
  \begin{equation}
    \lim_{r\rightarrow 2M} R = (r-2M)^{s_i}\;,
  \end{equation}
  then the exponents $s_i$ take the values $s_1=0$ and $s_2= 4\rho M$. In the sign conventions adopted here and the retarded time dependence $e^{\rho u}$, the $s_1$ branch is outgoing, whereas the $s_2$ branch is ingoing: \(e^{\rho u}(r-2M)^{4M\rho}\) has the near-horizon dependence \(e^{\rho v}\), where \(v=u+2r_{*}\). The QNM and quasi-bound problems both select the ingoing $s_2$ branch.
\item \textit{Irregular singular point at $r=\infty$}
  \\Finally, the asymptotic solutions of the master equation (\ref{eq:scalar_master_eq}) at infinity are 

  \begin{equation}
    \label{eq:scalar_behaviour_at_infinity}
R_{\infty}^{(\pm)}(r) \;\sim\;
\exp\!\big[(\rho \pm \kappa)\,r\big]\;
r^{\,\sigma_\pm}\;,
  \end{equation}
where

\begin{equation}
  \kappa = \sqrt{\rho^{2}+\scalaronmass^{2}} \qquad \text{and }\qquad \sigma_\pm = -1 \;\pm\;\frac{M}{\kappa}\,(\rho \pm \kappa)^{2}.
\end{equation}
The large-radius solutions define a formal two-branch asymptotic basis, whose physical identification depends on the sheet chosen for \(\kappa\). The convention for the QNM and quasi-bound problems is fixed in
Sec.~\ref{sec:mode_definitions_and_boundary_conditions}.

\end{itemize}

\section{Mode definitions, boundary conditions, and pole conditions}

\subsection{Scalar mode selections and asymptotic branches}
\label{sec:mode_definitions_and_boundary_conditions}
The master equations support several physically distinct mode selections. In what follows we focus on quasinormal modes and quasi-bound states.

\begin{itemize}
\item Quasinormal modes: For the separated frequency-domain problem, black hole quasinormal modes are traditionally defined by the horizon-admissible branch together with the outgoing large-radius branch. With the harmonic ansatz \eqref{eq:harmonic}, the damped modes correspond to $\Im(\omega)<0$, so that the amplitude decays in retarded time $u$. These conditions pick out a discrete set of complex frequencies, which appear as isolated poles of the retarded Green function in the lower half of the complex $\omega$-plane (see e.g. \cite{1986PhRvD..34..384L,Andersson:1996cm, Kokkotas:1999bd}).

\item Quasi-bound modes: For the massive scalar equation, the mass term changes the asymptotic structure at large $r$ and introduces an additional family of long-lived, spatially localised modes. In the separated frequency-domain problem, quasi-bound states are defined by the horizon-admissible branch together with the exponentially decaying large-radius branch in the evanescent regime. These are not true bound states because flux is absorbed by the horizon. Their frequencies are therefore complex with $\Im(\omega)<0$ for the Schwarzschild background \cite{PhysRevLett.109.081102,Sanchis-Gual:2016jst}.  
\end{itemize}

In this work, we use the term ``quasi-resonant limit''  for the weakly damped limit $\Im(\omega)\to 0^{-}$ of either spectal family.

For the asymptotic basis \(R_{\infty}^{(\pm)}\) in Eq.~\eqref{eq:scalar_behaviour_at_infinity}, we use
\begin{equation}
\kappa^2
=
\rho^2+\mu^2
=
\mu^2-\omega^2,
\qquad
\rho=-i\omega.
\label{eq:kappa_definition}
\end{equation}
The QNM sheet is fixed on the positive-frequency propagating axis by
\begin{equation}
\kappa
=
+i\sqrt{\omega^2-\mu^2},
\qquad
\omega>\mu,
\quad
\omega\in\mathbb{R},
\label{eq:kappa_qnm_sheet}
\end{equation}
for which \(R_{\infty}^{(+)}\) is outgoing. For the quasi-bound
problem, the branch is fixed in the real evanescent regime by
\begin{equation}
\kappa
=
+\sqrt{\mu^2-\omega^2},
\qquad
0<\omega<\mu,
\label{eq:kappa_qbs_sheet}
\end{equation}
and is continued so that \(\Re\kappa>0\); in this case \(R_{\infty}^{(-)}\) decays exponentially. Away from the real-frequency axis, these prescriptions are understood by analytic continuation on the selected sheet. Thus the QNM and quasi-bound problems select the \((+)\) and \((-)\) asymptotic branches, respectively.

%%%%%%%%%%%%%%%%%%%%%%%%%%%%%%%%%%%%%%%%%%%%%%%%%%%%%%%%%%%%%%%%%%%%%%%%%%%%%%%%%%%
\subsection{Scalar Wronskian and pole conditions}
\label{subsec:scalar_green_function}

Define $z=1-2M/r$ and let \(R_H(r,\rho)\) denote the horizon-admissible solution of the
scalar radial equation, normalised by
\[
R_H(r,\rho)=z^{4M\rho}[1+O(r-2M)],
\qquad
z\rightarrow0,
\]
and let \(R_\pm(r,\rho)\) denote the two large-radius solutions in
Eq.~\eqref{eq:scalar_behaviour_at_infinity}. The \(+\) solution is
evaluated on the outgoing QNM sheet, while the \(-\) solution is
evaluated on the decaying quasi-bound sheet, with the corresponding
\(\kappa\)-prescriptions fixed in
Sec.~\ref{sec:mode_definitions_and_boundary_conditions}.

For each branch, define the ordinary Wronskian
\begin{equation}
W_\pm(r,\rho)
=
R_H(r,\rho)R_\pm'(r,\rho)
-
R_H'(r,\rho)R_\pm(r,\rho).
\label{eq:ordinary_wronskian_green}
\end{equation}
Although \(W_\pm\) is not constant because the radial equation
contains a first-derivative term, Abel's identity implies that it
differs from an \(r\)-independent quantity
\(\mathcal W_\pm(\rho)\) only by a nonzero radial prefactor.
Consequently, \(W_\pm\) vanishes at one exterior radius if and only
if the corresponding endpoint solutions are linearly dependent.

In the standard radial Green-function construction, this Wronskian
appears in the denominator of the frequency-domain resolvent
\cite{1986PhRvD..34..384L}. Isolated zeros of
\(\mathcal W_\pm\) therefore give its poles. On the selected
\(\kappa\)-sheets, the scalar spectral conditions are
\begin{align}
\mathcal W_+(\rho_n)&=0,
&&\text{QNM condition},
\label{eq:qnm_wronskian_condition}
\\
\mathcal W_-(\rho_n)&=0,
&&\text{quasi-bound-state condition}.
\label{eq:qb_wronskian_condition}
\end{align}
The continued fractions introduced in
Sec.~\ref{subsec:cf_pole_condition} impose these same two-endpoint
matching conditions numerically.

The square-root branch points at \(\rho=\pm i\mu\) additionally give
branch-cut contributions, distinct from the isolated QNM and
quasi-bound poles.

%%%%%%%%%%%%%%%%%%%%%%%%%%%%%%%%%%%%%%%%%%%%%%%%%%%%%%%%%%%%%%%%%%%%%%%%%%%%%%%%%%%

\section{Frequency-domain formulation and numerical setup}
\label{sec:frequency_domain_formulation_and_numerical_setup}

\subsection{Continued-fraction representation of the pole condition}
\label{subsec:cf_pole_condition}

The Green-function discussion in \S\ref{subsec:scalar_green_function} shows that the scalar QNM and quasi-bound-state spectra are the zero sets of the modified Wronskians $\mathcal W_+(\rho)$ and $\mathcal W_-(\rho)$, respectively.  The role of the continued-fraction method is to impose the same two-endpoint matching condition numerically.

For either scalar branch, and similarly for the tensor sector in \S\ref{app:tensor_validation}, the Leaver expansion reduces the radial ODE to a three-term recurrence relation
\begin{subequations}
\label{eq:generic_three_term_recurrence}
\begin{align}
\alpha_0a_1+\beta_0a_0 &=0,\\
\alpha_ka_{k+1}+\beta_ka_k+\gamma_ka_{k-1}&=0, \qquad k\geq1.
\end{align}
\end{subequations}
The series is normalised at the horizon by choosing $a_0\neq0$. For generic $\rho$, the resulting coefficient sequence does not define the minimal solution as $k\to\infty$, and the corresponding series fails to satisfy the selected condition at the outer endpoint.  The eigenvalue condition is that the horizon-normalised sequence be minimal at infinity.

The corresponding infinite continued-fraction condition is
\begin{equation}
        \mathfrak F(\rho)
        \equiv
        \beta_0
        -
        \frac{\alpha_0\gamma_1}{
        \beta_1
        -
        \dfrac{\alpha_1\gamma_2}{
        \beta_2
        -
        \dfrac{\alpha_2\gamma_3}{
        \beta_3-\cdots}}}
        =0 .
\label{eq:generic_cf_condition}
\end{equation}
For the scalar sector, \(\mathfrak F_+\) and \(\mathfrak F_-\) denote the outgoing-QNM and decaying quasi-bound branches, respectively, with the \(\kappa\)-sheet fixed in Sec.~\ref{sec:mode_definitions_and_boundary_conditions}. In the infinite-continued-fraction limit,
\begin{equation}
        \mathfrak F_+(\rho)=0
        \quad\Longleftrightarrow\quad
        \mathcal W_+(\rho)=0 ,
\qquad \text{and} \qquad
        \mathfrak F_-(\rho)=0
        \quad\Longleftrightarrow\quad
        \mathcal W_-(\rho)=0 .
\end{equation}
This is an equivalence of zero sets, not an identity between the analytic functions themselves: $\mathfrak F_\pm$ depends on the normalization and series representation, whereas $\mathcal W_\pm$ is the Green-function denominator.  Both vanish when the horizon-admissible solution also satisfies the selected condition at infinity.

In practice, we solve finite, inverted versions of Eq.~\eqref{eq:generic_cf_condition}.  We denote the numerical root function
by
\begin{equation}
  \label{eq:finite_cf_root_function}
  F_{\chi,n}^{(N)}(\rho),\qquad \chi \in \{+,- \}\;,
\end{equation}
where $\chi$ labels the scalar branch and $n$ labels the inversion used to target a given overtone, with $N$ being the truncation index.  At finite $N$ the
equation
\begin{equation}
  \label{eq:finite_cf_root_condition}
  F_{\chi,n}^{(N)}(\rho)=0\;,
\end{equation}
is only an approximation to the infinite pole condition.  Accepted roots must therefore remain stable as
\(N\) is increased. Roots that drift or disappear under this test are
treated as finite-truncation artefacts rather than poles of the
resolvent.

  \subsection{Scalar (spin-0) sector}
\label{sec:scalar_freq}

The calculations of the recurrence relation for the QNM and quasi-bound spectra proceeds in similar ways, with the resulting coefficients taking the same structure. Thus, to avoid repetitions, let
\[
\varepsilon=\begin{cases}
+1, & \text{outgoing QNM branch},\\
-1, & \text{decaying quasi-bound branch},
\end{cases}
\]
with the corresponding \(\kappa\)-sheet fixed in Sec.~\ref{sec:mode_definitions_and_boundary_conditions}. In addition, define
\begin{equation}
  \label{eq:branch_dependent_parameters}
\lambda_\varepsilon=\rho+\varepsilon\kappa,\qquad\sigma_\varepsilon=-1+\varepsilon\frac{M}{\kappa}\lambda_\varepsilon^2.
\end{equation}
Now, using the near-horizon and asymptotic forms outlined in \S\ref{sec:scalar_perturbations}, the two spectral problems can be represented by the common Frobenius
ansatz
\begin{equation}
  \label{eq:scalar_branch_ansatz}
R_\varepsilon(r)=e^{\lambda_\varepsilon r}r^{\sigma_\varepsilon}z^{4M\rho}\sum_{n=0}^{\infty}a_nz^n\;, \qquad z=1-\frac{2M}{r}.
\end{equation}
with the particular exponents in the prefactors ensuring the desired behaviors at the horizon and null infinity. Substituting this ansatz into the massive scalar master equation yields a three–term recurrence relation
\begin{subequations}
  \label{eq:scalar_branch_recurrence}
\begin{align}
%% \alpha_n&=(n+1)(n+1-4M\rho),\\
%% \beta_n^{(\varepsilon)}&=-2n^2+\left(4M\lambda_\varepsilon+2\sigma_\varepsilon\right)n+\sigma_\varepsilon-4M\sigma_\varepsilon\lambda_\varepsilon-2M\lambda_\varepsilon-\ell(\ell+1),\\
%% \gamma_n^{(\varepsilon)}&=(n-1-\sigma_\varepsilon)^2.
\alpha_n&=(n+1)(n+1+4M\rho),\\
\beta_n^{(\varepsilon)}&=-2(n+4M\rho)^2+\left(4M\lambda_\varepsilon+2\sigma_\varepsilon\right)(n+4M\rho) \nonumber\\
&\mathrel{\phantom{=}}+\sigma_\varepsilon-4M\sigma_\varepsilon\lambda_\varepsilon-2M\lambda_\varepsilon-\ell(\ell+1),\\
\gamma_n^{(\varepsilon)}&=(n+4M\rho-1-\sigma_\varepsilon)^2.
\end{align}
\end{subequations}
In the notation of \S\ref{subsec:cf_pole_condition}, the QNM and quasi-bound frequencies are the roots of \({F^{(N)}_{+,n}(\rho)=0}\) and \({F^{(N)}_{-,n}(\rho)=0}\), respectively. In order to apply Nollert's improvement in this case we perform a large–$N$ approximation of the recurrence and expand the remainder term in inverse fractional powers of $N$
\begin{equation}
R_N^{(\varepsilon)}\sim C_0^{(\varepsilon)}+\frac{C_1^{(\varepsilon)}}{\sqrt{N}}+\frac{C_2^{(\varepsilon)}}{N}+\mathcal{O}\!\left(N^{-3/2}\right),
\label{eq:RN_scalar_general}
\end{equation}
The coefficients $C^{\varepsilon}_i$ are determined by the large-\(k\) behaviour of \(\alpha_k\), \(\beta_k^{(\varepsilon)}\), and \(\gamma_k^{(\varepsilon)}\). For the recurrence \eqref{eq:scalar_branch_recurrence}, we obtain
\begin{equation}
C_0^{(\varepsilon)}
=
-1\;,
\qquad
\left(C_1^{(\varepsilon)}\right)^2
=
-4\varepsilon M\kappa\;,
\qquad
C_2^{(\varepsilon)}
=
\sigma_\varepsilon
+
\frac{7}{4}
-
2M\lambda_{-\varepsilon}\;.
\label{eq:scalar_branch_tail_coefficients}
\end{equation}
Choosing the sign of \(C_1^{(\varepsilon)}\) so that the large-\(k\) recurrence selects the minimal solution, the expressions for two remainders can be written explicitly as
\begin{subequations}
\label{eq:scalar_branch_remainders}
\begin{align}
R_N^{(+)}
&\simeq
-1
+
\frac{2\sqrt{-M\kappa}}{\sqrt N}
+
\frac{\sigma_+ + 7/4 - 2M\lambda_-}{N}\;,
\\
R_N^{(-)}
&\simeq
-1
+
\frac{2\sqrt{M\kappa}}{\sqrt N}
+
\frac{\sigma_- + 7/4 - 2M\lambda_+}{N}\;.
\end{align}
\end{subequations}
Here \(\kappa\) is evaluated on the sheet fixed in
Sec.~\ref{sec:mode_definitions_and_boundary_conditions}. The
remaining sign ambiguity in \(C_1^{(\varepsilon)}\) is fixed by the
minimal-solution condition and continued smoothly along each tracked
root branch.

\subsection{Numerical implementation}

For a fixed multipole $\ell$, branch/sector label $\chi$, overtone index $n$, and truncation index $N$, we solve the numerical root problem \eqref{eq:finite_cf_root_function}. Here $\chi=+$ denotes the scalar QNM branch, $\chi=-$ denotes the scalar quasi-bound branch, and $\chi=\mathrm{T}$ denotes the tensor QNM branch with the tensor recurrence coefficients used in the consistency check of
Appendix~\ref{app:tensor_validation}.  The index $n$ labels the inversion used to target the $n$th overtone, while $N$ is the truncation index used in the backward evaluation of the continued fraction.  When no confusion can arise, we suppress the labels $\chi$ and $N$ and write simply $F_n(\rho)$.

As in \cite{Mongwane:2024vao}, the root finding scheme used in this work requires initial guesses for each $n$. We construct initial data as follows:
\begin{itemize}
\item For the first two overtones $n=1$ and $n=2$, we initialise the search using an asymptotic expansion for the QNM frequency in inverse powers of $L=\ell+1/2$ (evaluated at the desired $\ell$)
\begin{equation}
\omega_{\ell n} = \varpi^{(n)}_{-1}L+\varpi^{(n)}_{0}+\varpi^{(n)}_{1}L^{-1}+\varpi^{(n)}_{2}L^{-2}+\cdots
\end{equation}
where the expressions of the $\varpi_{i}^{(n)}$ are given in \cite{Dolan:2009nk}.
\item For higher overtones ($n>2$), we use mode tracking by extrapolation from the previous two roots. If $\rho_{n-2}$ and $\rho_{n-1}$ are known (at fixed $(\ell,m)$ and fixed external parameters), then we take
\begin{equation}
\label{eq:linear_extrap_guess}
\rho^{(0)}_{n}=\rho_{n-1}+\bigl(\rho_{n-1}-\rho_{n-2}\bigr),
\end{equation}
i.e.\ linear extrapolation in the complex plane.
\end{itemize}
Finally, the continued fractions entering $F_n(\rho)=0$ are evaluated by backward iteration: starting from a truncated index $N$, we initialise the tail using the asymptotic remainder \eqref{eq:tail_tensor} in the tensor case and \eqref{eq:scalar_branch_remainders} in the scalar case, and then iterate downward to obtain the required finite value. In this setting, backward evaluation is numerically robust, so a Lentz-type algorithm is not required.

\section{Time-domain formulation and numerical setup}
\label{sec:time_domain_formulations_and_numerical_setup}

\subsection{Scalar characteristic evolution scheme}
\label{sec:characteristic_evolution_scheme}

In the time-domain, the scalaron master equation is given by \eqref{eq:scalar_master_eq}, with factors of $\rho$ converted back to differentiation with respect to $u$ according to \ref{eq:dictionary}. In particular, we have
\begin{equation}
  \label{eq:scalar_master_eq_td}
  r\left(r-2M\right)R_{,rr}-2r^2R_{,ur}-2\left(-r +M\right)R_{,r}-2rR_{,u}-\left[\ell(\ell+1)+\scalaronmass^2r^2\right]R=0\;.
\end{equation}
Introducing $\psi\equiv R_{,u}$, this can be conveniently written as an evolution equation plus hypersurface equation, a configuration which is ideally suited for evolution via a characteristic scheme,
\begin{subequations}
\label{eq:scalar_td_system}
\begin{align}
R_{,u}
&=
\psi\;,
\\
\psi_{,r}
+
\frac{\psi}{r}
&=
\frac{r-2M}{2r}R_{,rr}
+
\frac{r-M}{r^2}R_{,r}
-
\frac{\ell(\ell+1)+\mu^2r^2}{2r^2}R\;.
\end{align}
\end{subequations}
In order to place the inner worldtube close to the future horizon, we introduce
\begin{equation}
x
=
\log\left(\frac{r}{2M}-1\right),
\qquad
q\equiv r-2M=2Me^x,
\qquad
\frac{dr}{dx}=q,
\label{eq:td_log_coordinate}
\end{equation}
and define \(y=r\psi\). The radial hypersurface equation then takes the regular integrated form
\begin{equation}
y_{,x}=F_R,
\label{eq:td_integrated_hypersurface}
\end{equation}
where
\begin{equation}
F_R=\frac{1}{2}R_{,xx}+\frac{q}{2r}R_{,x}-\frac{q}{2r}\left[\ell(\ell+1)+\mu^2r^2\right]R.
\label{eq:td_scalar_source}
\end{equation}
The evolution equation is advanced in $u$ using the classic explicit fourth order Runge-Kutta method, and the hypersurface equation is integrated outward on the uniform logarithmic \(x\)-grid using a fourth-order Euler--Maclaurin-corrected trapezoidal rule, with fourth-order finite differences used for the spatial derivatives. The spatial derivatives entering \(F_R\) are evaluated using fourth-order finite differences, with one-sided fourth-order closures at the boundaries. Details of the radial discretization are given in Appendix~\ref{app:characteristic_discretization}.

\subsection{Initial data and boundary treatment}
\label{sec:td_initial_boundary}

The evolution is initialised by prescribing the master variable \(P\) on
the initial null hypersurface.  We use Gaussian data
\begin{equation}
  R(u=0,r)
  =
  A_0
  \exp\left[-\frac{(r-r_0)^2}{2\sigma^2}\right],
  \label{eq:td_gaussian_initial_data}
\end{equation}
with \(A_0=1\).  Since the system is linear, the overall amplitude does
not affect the extracted frequencies.

The inner worldtube is placed at fixed \(x=x_{\min}\), with
\(x\) defined in Eq.~\eqref{eq:td_log_coordinate}.  For the production
scalar runs we typically use \(x_{\min}=-35\), corresponding to
\(r_{\min}-2M=2M e^{-35}\). 
The inner-worldtube condition is obtained from the leading ingoing
behaviour near the future horizon. The ingoing radial null curves
of the background metric~\eqref{eq:background} are labelled by
\[
    v=u+2r_*,
    \qquad
    r_*=r+2M\log\left(\frac{r}{2M}-1\right).
\]
With the logarithmic coordinate~\eqref{eq:td_log_coordinate}, \(dr_*/dx=r\).
Writing the scalar perturbation as \footnote{For brevity, we use \(\mathcal R\) to denote the same scalar field expressed in ingoing coordinates, i.e. \(\mathcal{R} \) is not a rescaled master variable or an additional evolved quantity}
\(R(u,r)=\mathcal{R}(v,r)\), where \(\mathcal{R}\) is regular
at the future horizon, the chain rule gives
\[
    R_{,u}-\frac{1}{2r}R_{,x}
    =
    -\frac{r-2M}{2r}
    \left.\frac{\partial\mathcal{R}}{\partial r}\right|_v.
\]
For bounded radial derivatives in the ingoing coordinates, the
right-hand side vanishes as \(r\to2M\). Neglecting this
horizon-suppressed contribution at the inner worldtube therefore
supplies the boundary value required by the hypersurface solve:
\begin{equation}
    \psi(u,x_{\min})
    =
    \frac{R_{,x}(u,x_{\min})}{2r_{\min}},
    \qquad
    y(u,x_{\min})
    =
    \frac{1}{2}R_{,x}(u,x_{\min}).
    \label{eq:td_inner_bc}
\end{equation}
This is a leading near-horizon ingoing approximation at a finite
worldtube, rather than an exact absorbing condition at arbitrary
radius. Sensitivity to the worldtube location is tested in
Table~\ref{tab:td_robustness}.

No independent outer-boundary data is imposed for \(\psi\), which is obtained by radial integration from the inner worldtube.  The outer radius
\(r_{\max}\) is instead chosen so that the waveform interval used for
convergence and mode extraction lies before any expected finite-domain
contamination.  We monitor this using the estimate
\begin{equation}
  u_{\rm refl}
  \simeq
  2\left[
    r_*(r_{\max})-r_*(r_{\rm obs})
  \right],
  \qquad
  r_*=r+2M\log\left(\frac{r}{2M}-1\right).
  \label{eq:td_reflection_time}
\end{equation}
The dependence on \(r_{\max}\) and on the extraction radius \(r_{\rm obs}\) is included in the robustness tests presented in Table~\ref{tab:td_robustness}.

This boundary choice defines the characteristic initial-value problem. The finite-radius waveform resulting from this initial--worldtube problem is analysed in \S\ref{sec:mode_content_identification}.

\subsection{Mode content and pole identification}
\label{sec:mode_content_identification}

\begin{table}[t]
\centering
\small
\setlength{\tabcolsep}{5pt}
\renewcommand{\arraystretch}{1.15}
\begin{tabularx}{\linewidth}{
@{}
>{\raggedright\arraybackslash}p{0.29\linewidth}
>{\raggedright\arraybackslash}X
@{}
}
  %\toprule
  \hline
\textbf{Calculation}
&
\textbf{Selection or interpretation}
\\
  \hline%\midrule
%\addlinespace
Frequency-domain QNM
&
Regularity at the future horizon and the outgoing large-radius
branch select the discrete QNM spectrum.
\\
  \hline
Frequency-domain quasi-bound state
&
Regularity at the future horizon and the exponentially decaying
large-radius branch select the discrete quasi-bound spectrum.
\\
  \hline
Characteristic evolution
&
The inner-worldtube boundary value required by the radial
hypersurface solve on each null slice. No independent outgoing or
decaying spectral branch is imposed at \(r_{\max}\), so no spectral
family is selected a priori.
\\
  \hline
Time-domain pole identification
&
Stable poles are identified a posteriori under variations of fitting
window, pencil rank, and decimation. Continued-fraction values are
introduced only after this stability selection.
\\

  \hline%\bottomrule
\end{tabularx}
\caption{Frequency-domain mode selection and time-domain mode
identification.}
\label{tab:mode_selection}
\end{table}
The frequency-domain and characteristic calculations implement different mode-selection procedures. Table~\ref{tab:mode_selection} summarises this distinction and the a posteriori identification used in the time-domain analysis. We note that the frequency-domain calculations are spectral boundary-value problems, whereas the characteristic calculation produces a finite-radius initial-value response. The resulting waveform may contain prompt, QNM, quasi-bound, branch-cut, and other non-modal components \cite{1986PhRvD..34..384L,Andersson:1996cm,Bhattacharyya:2020zkc}. Their relative excitation depends on the initial data, but their complex frequencies are properties of the underlying evolution equation. Candidate modes are therefore identified through their persistence under changes of the extraction procedure rather than from waveform morphology or from a single fitting interval.

For each fitting window, the mean-subtracted waveform is approximated by
a finite sum of damped exponentials,
\begin{equation}
  R(u_j,r_{\rm obs})-\overline{R}_{W}
  \simeq
  \sum_{k=1}^{p} c_k e^{\lambda_k u_j},
  \label{eq:matrix_pencil_model}
\end{equation}
where \(p\) is the retained SVD rank, which sets the exponential model order for that fit. The complex exponents
\(\lambda_k\) are obtained using an SVD-truncated matrix-pencil method \cite{Berti:2007dg,Sarkar:1995,Hua:1990}. Matrix-pencil and related Prony-type estimators have been applied to black-hole ringdown and scalar-field time-domain waveforms, e.g. \cite{Berti:2007dg,Bhattacharyya:2020zkc}.  More generally,
varying the fitting interval and model order, and retaining only
components that persist under these changes, is a common strategy for
robust ringdown-mode identification \cite{Cheung:2023vki}. In this work, we augment the standard matrix-pencil estimator with residual filtering, graph-based clustering, and window/rank/decimation diversity tests. For a real-valued waveform, oscillatory contributions occur in approximately complex-conjugate pairs.  We define
\begin{equation}
  \gamma_k=-\Re(\lambda_k)>0,
  \qquad
  \Omega_k=\left|\Im(\lambda_k)\right|,
\end{equation}
and report the corresponding complex frequency as
\begin{equation}
  2M\omega_{\rm TD}
  =
  2M\Omega_k-i\,2M\gamma_k .
  \label{eq:td_complex_frequency}
\end{equation}

The extraction is repeated over a family of fitting-window start times,
window lengths, matrix-pencil ranks, and decimation factors.  This
produces an ensemble of candidate poles rather than a single
fit-dependent frequency.  Candidate poles are first required to satisfy
\(\gamma_k>0\) and to lie within prescribed intervals of
\(\Re(2M\omega)\) and \(-\Im(2M\omega)\).  For each fit we also compute
the relative reconstruction residual
\begin{equation}
  \epsilon_{\rm rec}
  =
  \frac{\|R_W-\widehat{R}_W\|_2}
       {\|R_W\|_2},
  \label{eq:matrix_pencil_residual}
\end{equation}
where \(R_W\) denotes the mean-subtracted waveform in the fitting
window.  An adaptive residual prefilter removes poles belonging to fits
whose residual exceeds the selected residual quantile or absolute
threshold.

The remaining poles are grouped using an
\(\epsilon\)-neighbour connected-component algorithm in the plane
\begin{equation}
  \left(
    \Re(2M\omega),
    -\Im(2M\omega)
  \right).
\end{equation}
A component is retained only if it contains a minimum number of poles
and is supported by sufficiently diverse choices of window start,
window length, matrix-pencil rank, and decimation factor.  This
diversity requirement reduces the likelihood that a pole appearing only
for one special fit is identified as a physical mode.  Surviving
clusters are ranked using their population, reconstruction residual,
and spread in the complex-frequency plane.

For an accepted cluster, we define
\begin{equation}
  \widetilde{\Omega}_{2M}
  =
  {\rm median}\!\left[\Re(2M\omega_k)\right],
  \qquad
  \widetilde{\Gamma}_{2M}
  =
  {\rm median}\!\left[-\Im(2M\omega_k)\right],
\end{equation}
and report
\begin{equation}
  2M\omega_{\rm TD}
  =
  \widetilde{\Omega}_{2M}
  -
  i\widetilde{\Gamma}_{2M}.
\end{equation}
The componentwise spreads are quantified by
\begin{align}
  {\rm MAD}_{\Omega}
  &=
  {\rm median}\!
  \left(
    \left|
      \Re(2M\omega_k)-\widetilde{\Omega}_{2M}
    \right|
  \right),
  \\
  {\rm MAD}_{\Gamma}
  &=
  {\rm median}\!
  \left(
    \left|
      -\Im(2M\omega_k)-\widetilde{\Gamma}_{2M}
    \right|
  \right).
  \label{eq:td_cluster_mads}
\end{align}
The componentwise MADs quantify stability under variations of the extraction procedure and are not interpreted as formal statistical uncertainties.

The continued-fraction frequencies do not enter the matrix-pencil solve, candidate-pole filters, clustering algorithm, diversity criteria, or cluster ranking.  They are added only after the time-domain cluster has been determined, as independent reference values for the subsequent comparison.

Lastly, we note that the stability tests used here concern the numerical persistence of a pole extracted from a fixed evolution equation. They should not be confused with pseudospectral stability, which concerns the displacement of QNM eigenvalues under perturbations of the underlying radial operator \cite{Jaramillo:2020tuu,Cardoso:2024mrw}. A pseudospectral analysis of the massive scalaron operator lies outside the scope of the current work.

\section{Results}
\label{sec:results}

We focus below on the scalaron QNM and quasi-bound spectra and on their recovery both in the frequency domain and characteristic time-domain evolution. Before presenting the scalar spectra, we emphasise that the curves below should be interpreted as universal massive-scalar Schwarzschild spectra parameterised by \(\mu M\), rather than as spectra tied to a particular polynomial choice of \(f(R)\) (See the discussion in \S\ref{sec:fr_perturbative_reduction}).

\subsection{Scalar quasinormal modes and frequency-domain validation}
\label{sec:scalar_qnm_and_frequency_domain_validation}

We first validate the scalar continued-fraction implementation against selected frequency-domain results from the literature. The benchmark set in Table~\ref{tab:scalar_cf_benchmarks} is chosen to
match the modes for which characteristic time-domain extractions are reported in Table~\ref{tab:scalar_td_benchmarks}. This enables a three-way comparison between values available in the literature, the continued fraction calculation in this work, and the characteristic time-domain results. The selected cases comprise a massless multipole test for $\ell=0,\ldots,3$, a low-mass multipole test at $\mu M=0.05$ for $\ell=0,\ldots,4$, and finite-mass sequences for $\ell=1,2$ at $\mu M=0.10,0.15,$ and $0.20$. Since no single reference tabulates all these cases in one table, we compile the literature benchmarks from several complementary references \cite{Cavalcante:2021scq,Alves:2025erh,Dolan:2007mj,Konoplya:2017tvu,Konoplya:2026rjh}. The conventions adopted for the reference values (along with any required unit conversions) are detailed in Appendix~\ref{app:frequencies_from_literature}. For each benchmark, we quantify the agreement using the absolute complex-frequency difference
\begin{equation}
  \label{eq:delta_cf_ref}
    \left|\Delta_{\mathrm{CF-ref}}\right|
    =
    \left|
        2M\omega_{\mathrm{CF}}^{\mathrm{this}}
        -
        2M\omega_{\mathrm{ref}}
    \right|.
\end{equation}
When evaluating Eq.~\eqref{eq:delta_cf_ref}, we use the full numerical precision available in each reference, rather than the rounded values displayed in the table.

Table~\ref{tab:scalar_cf_benchmarks} shows that our continued-fraction results reproduce the literature benchmarks across the massless, small-mass, and moderate-mass regimes.  The differences are generally at or below the $10^{-6}$ level, to the extent allowed by the precision and numerical truncation of the values as reported in each reference.  The comparatively larger value for the $\ell=0$, $\mu M=0.05$ benchmark is discussed in Appendix~\ref{app:frequencies_from_literature}. Having established this agreement, we now use the continued-fraction implementation in the following to track the scalar QNM branches as $\mu M$ is varied. 

\begin{figure}[!h]
      \centering
      \includegraphics[width=0.45\textwidth]{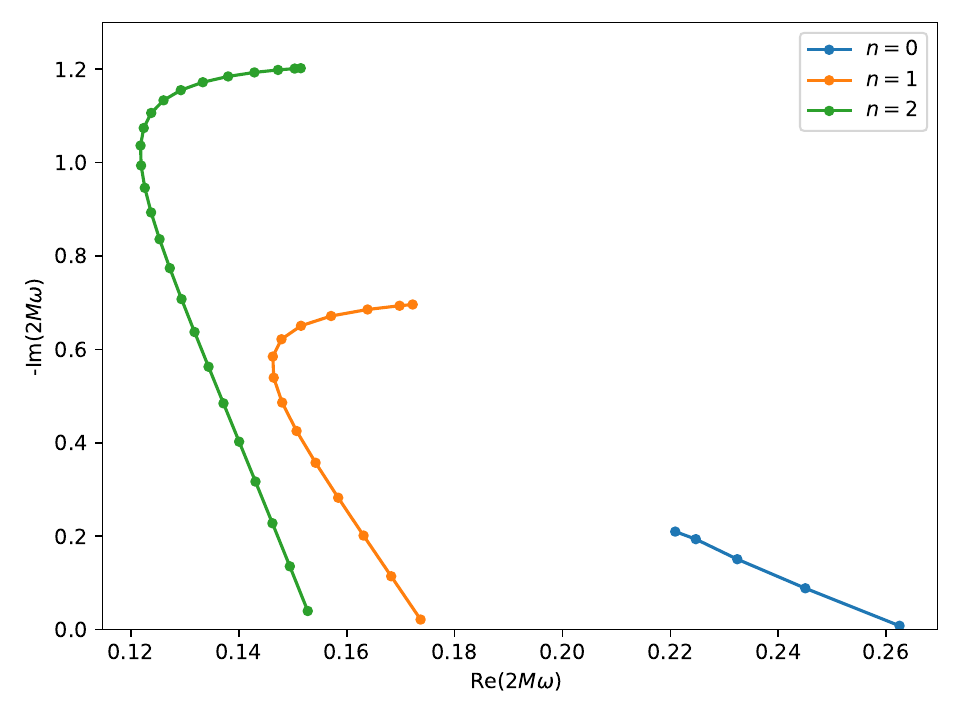}
      \includegraphics[width=0.45\textwidth]{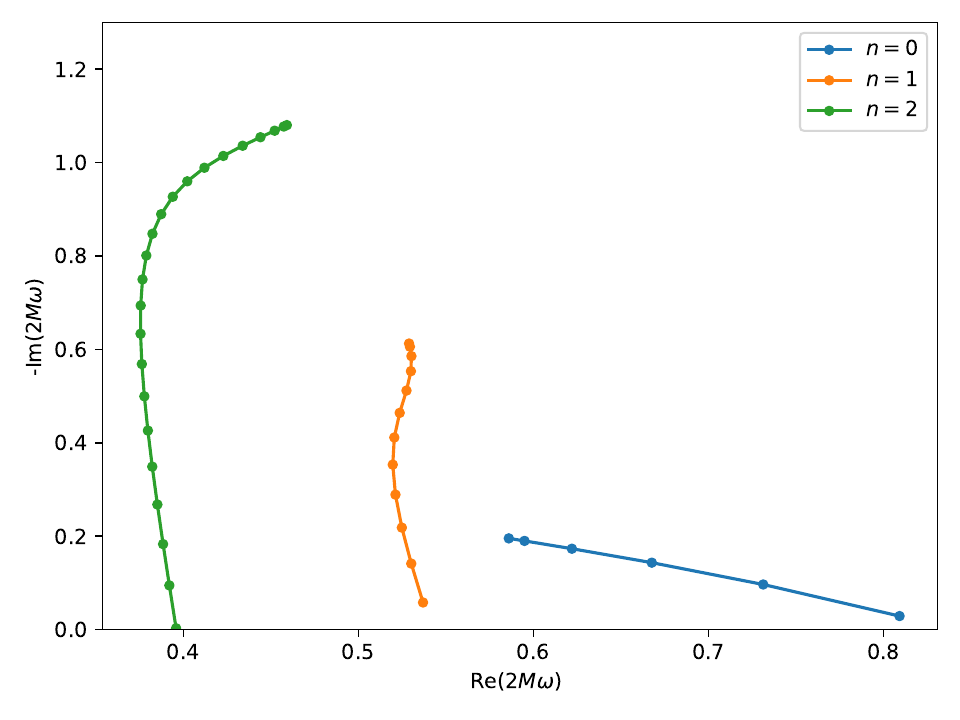}
 \caption{Root locus of massive-scalar QNMs for $\ell=0$ (left panel) and $\ell=1$ (right panel), computed using Leaver's continued-fraction method. The branches correspond to the fundamental mode and first two overtones. The markers indicate the values as $\mu M$ is varied from $\mu M=0$ in increments of $0.1$; curves connect successive points along each branch.}
      \label{fig:mass_scan_l0}
    \end{figure}

\begin{figure}[!h]
      \centering
      \includegraphics[width=0.45\textwidth]{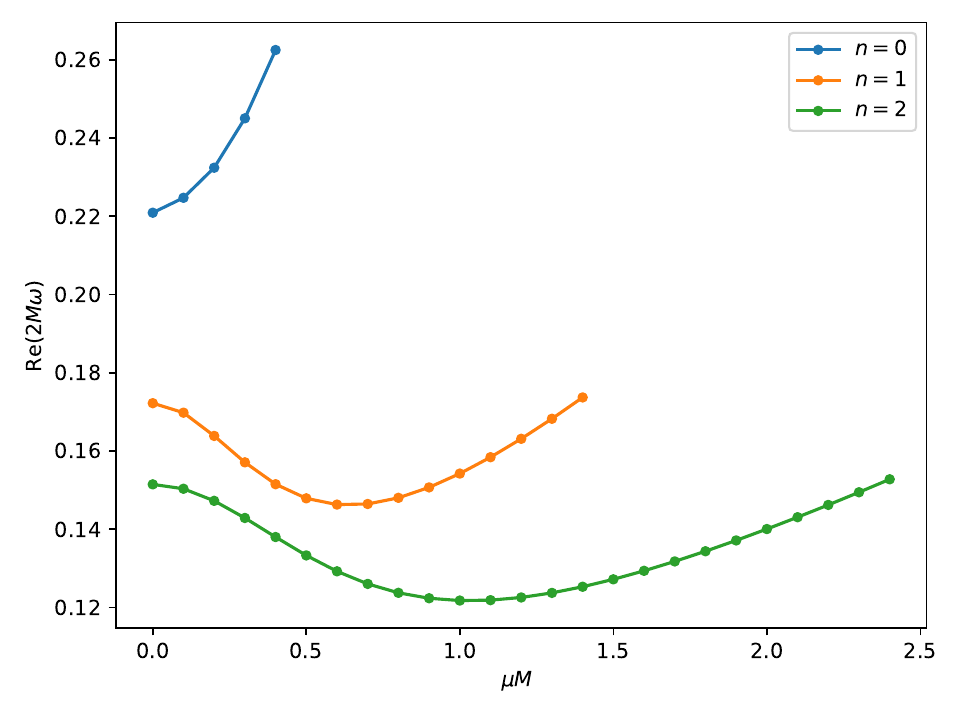}
      \includegraphics[width=0.45\textwidth]{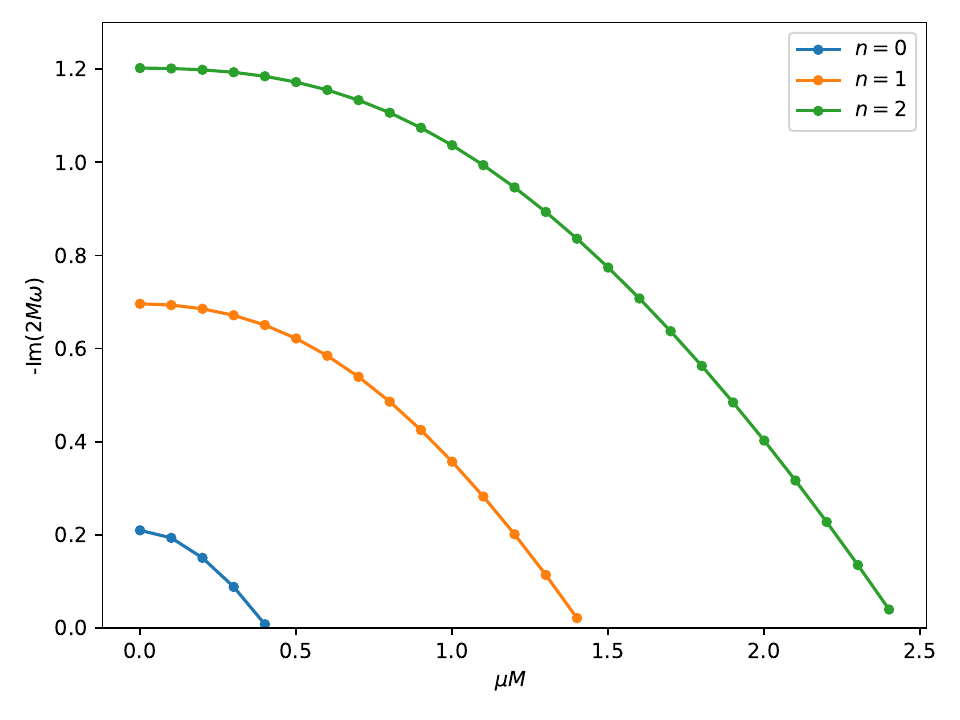}
      \caption{Mass dependence of the fundamental mode \(n=0\) and first two overtones \(n=1,2\) of the scalaron quasinormal spectrum for \(\ell=0\), computed using Leaver's continued-fraction method. The left panel shows the oscillation frequency \(\re(2M\omega)\), and the right panel shows the damping magnitude \(-\im(2M\omega)\), as functions of \(\mu M\). Markers denote computed frequencies, and lines connect successive samples. These are the \(\ell=0\) branches shown in Figure \ref{fig:mass_scan_l0}}
      \label{overtones_as_function_of_field_mass0}
    \end{figure}

\begin{figure}[!h]
      \centering
      \includegraphics[width=0.45\textwidth]{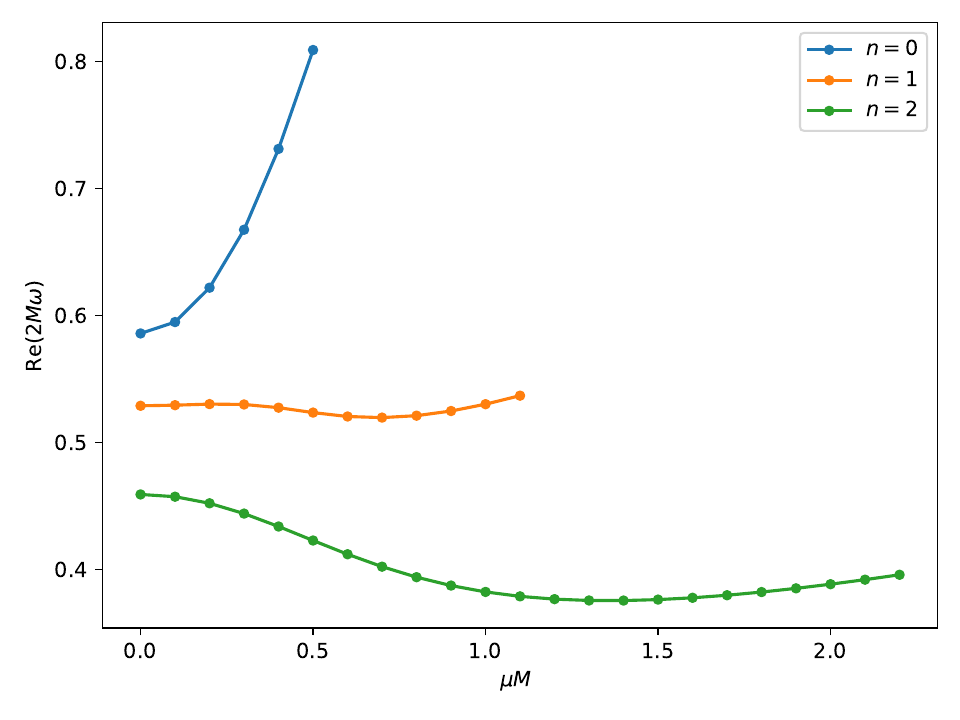}
      \includegraphics[width=0.45\textwidth]{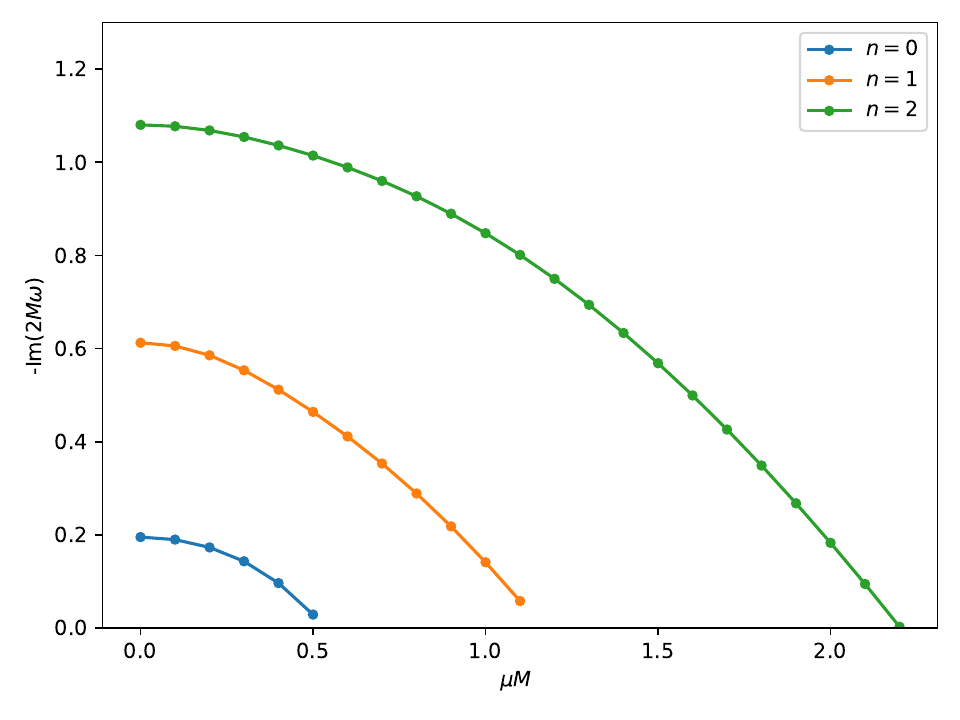}
      \caption{Mass dependence of the fundamental mode \(n=0\) and first two overtones \(n=1,2\) of the scalaron quasinormal spectrum for \(\ell=1\), computed using Leaver's continued-fraction method. The left panel shows the oscillation frequency \(\re(2M\omega)\), and the right panel shows the damping magnitude \(-\im(2M\omega)\), as functions of \(\mu M\). Markers denote computed frequencies, and lines connect successive samples. These are the \(\ell=0\) branches shown in Figure \ref{fig:mass_scan_l0}}
      \label{overtones_as_function_of_field_mass1}
    \end{figure}

\begin{table}[t]
    \centering

    \begingroup
    \footnotesize
    \setlength{\tabcolsep}{4.0pt}
    \renewcommand{\arraystretch}{1.12}

    \begin{tabular}{|c|c|c|c|c|}
              \hline
        \toprule
        $(\ell,\mu M)$
        & $2M\omega_{\mathrm{CF}}^{\mathrm{this}}$
        & $2M\omega_{\mathrm{ref}}$
        & Ref.
        & $\left|\Delta_{\mathrm{CF-ref}}\right|$
        \\
        \hline
        \midrule
        \multicolumn{5}{|c|}{
            \itshape Massless Schwarzschild normalization benchmark
        }
        \\
%        \addlinespace[1pt]
        \hline
        $(0,0.00)$
        & $0.220910-0.209791\,i$
        & $0.220910-0.209791\,i$
        & \cite{Cavalcante:2021scq}
        & $8.72104\times10^{-13}$
        \\

        $(1,0.00)$
        & $0.585872-0.195320\,i$
        & $0.585872-0.195320\,i$
        & \cite{Cavalcante:2021scq}
        & $1.01566\times10^{-12}$
        \\

        $(2,0.00)$
        & $0.967288-0.193518\,i$
        & $0.967288-0.193518\,i$
        & \cite{Cavalcante:2021scq}
        & $8.13174\times10^{-13}$
        \\

        $(3,0.00)$
        & $1.350732-0.192999\,i$
        & $1.350732-0.192999\,i$
        & \cite{Cavalcante:2021scq}
        & $7.60237\times10^{-13}$
        \\
        \hline
        \midrule

        \multicolumn{5}{|c|}{
            \itshape Low-mass multipole benchmark
        }
        \\
%        \addlinespace[1pt]
        \hline
        $(0,0.05)$
        & $0.221990-0.205692\,i$
        & $0.221976-0.205684\,i$
        & \cite{Alves:2025erh}
        & $1.55706\times10^{-5}$
        \\

        $(1,0.05)$
        & $0.588109-0.193976\,i$
        & $0.588108-0.193976\,i$
        & \cite{Alves:2025erh}
        & $6.32838\times10^{-7}$
        \\

        $(2,0.05)$
        & $0.968866-0.192976\,i$
        & $0.968866-0.192976\,i$
        & \cite{Alves:2025erh}
        & $3.53999\times10^{-7}$
        \\

        $(3,0.05)$
        & $1.351912-0.192714\,i$
        & $1.351912-0.192714\,i$
        & \cite{Alves:2025erh}
        & $3.13370\times10^{-7}$
        \\

        $(4,0.05)$
        & $1.735766-0.192609\,i$
        & $1.735766-0.192609\,i$
        & \cite{Alves:2025erh}
        & $2.77415\times10^{-7}$
        \\
        \hline
  %      \midrule

        \multicolumn{5}{|c|}{
          %\itshape Selected finite-mass benchmarks
              \itshape Finite-mass benchmarks matched to the time-domain study
        }
        \\
        \hline
 %       \addlinespace[1pt]

        $(1,0.10)$
        & $0.594831-0.189914\,i$
        & $0.594832-0.189914\,i$
        & \cite{Dolan:2007mj}
        & $6.93318\times10^{-7}$
        \\
        $(2,0.10)$
        & $0.973608-0.191349\,i$
        & $0.973608-0.191349\,i$
        & \cite{Konoplya:2017tvu}
        & $5.22153\times10^{-7}$
        \\
        $(1,0.15)$
        & $0.606080-0.183041\,i$
        & $0.606080-0.183041\,i$
        & \cite{Konoplya:2026rjh}
        & $3.84668\times10^{-11}$
        \\

        $(2,0.15)$
        & $0.981527-0.188624\,i$
        & $0.981527-0.188624\,i$
        & \cite{Konoplya:2017tvu}
        & $5.88887\times10^{-7}$
        \\

        $(1,0.20)$
        & $0.621914-0.173187\,i$
        & $0.621914-0.173186\,i$
        & \cite{Dolan:2007mj}
        & $5.99891\times10^{-7}$

        \\

        $(2,0.20)$
        & $0.992653-0.184778\,i$
        & $0.992653-0.184778\,i$
        & \cite{Konoplya:2017tvu}
        & $3.94598\times10^{-7}$
        \\
%        \bottomrule
        \hline
    \end{tabular}
        \caption{
    Frequency-domain benchmark of the fundamental ($n=0$) scalar
    Schwarzschild quasinormal modes. All frequencies are displayed in
    the convention
    $2M\omega=\Re(2M\omega)-i[-\Im(2M\omega)]$
    and are rounded to six decimal places. The differences
    $\left|\Delta_{\mathrm{CF-ref}}\right|$ are evaluated using the
    full numerical precision available in this work and in the
    corresponding reference, and are displayed to six significant
    figures.
    }
    \label{tab:scalar_cf_benchmarks}
    \endgroup
\end{table}
Along the fundamental branches shown, when the mass of the scalar field increases, the oscillation frequency increases, while the damping decreases. The real parts of the higher overtones are not monotonic over the full range displayed, whereas their damping follows the same general trend. This is consistent with the earlier results on scalar perturbations of the Schwarzschild black hole \cite{Simone_1992,Ohashi:2004wr,Konoplya:2004wg}. In Fig~\ref{fig:mass_scan_l0}, we show the root locus of the fundamental mode and first two overtones $n=0,1,2$ for $\ell=0$ (left panel) and $\ell=1$ (right panel) as $\mu M$ increases from $\mu M=0$ in steps of $0.1$. For both $\ell=0$ and $\ell=1$, each overtone traces a smooth curve on the complex plane, starting from their massless values at $\mu M=0$ moving toward a quasi-resonant regime $\im(2M\omega)\to 0^{-}$ as $\mu M$ is increased. In fact, this general behavior is robust, and persists across different multipoles \cite{Alves:2025erh}. Overall, lower overtones approach this regime at lower masses compared to the higher overtones. This can more clearly be seen from the right panels of Figs \ref{overtones_as_function_of_field_mass0} ($\ell=0$) and \ref{overtones_as_function_of_field_mass1} (for $\ell=1$). It is important to note that this vanishing of the imaginary part does not lead to an analogous vanishing of the real part as seen from the right panels of Figs~\ref{overtones_as_function_of_field_mass0} and \ref{overtones_as_function_of_field_mass1} which shows the corresponding real part of the modes. In other words, increasing $\mu M$ predominantly affects damping rather than oscillation, i.e. as $\im(2M\omega)\to 0^{-}$, the decay time $\tau\sim 1/|\im(2M\omega)|$ becomes arbitrarily large, thus yielding modes that are, essentially, arbitrarily long-lived while retaining non-zero oscillation frequency, which is determined by $\re(2M\omega)$.

    \subsection{Scalar quasi-bound states and frequency-domain validation}
\begin{figure}[!h]
      \centering
      \includegraphics[width=0.65\textwidth]{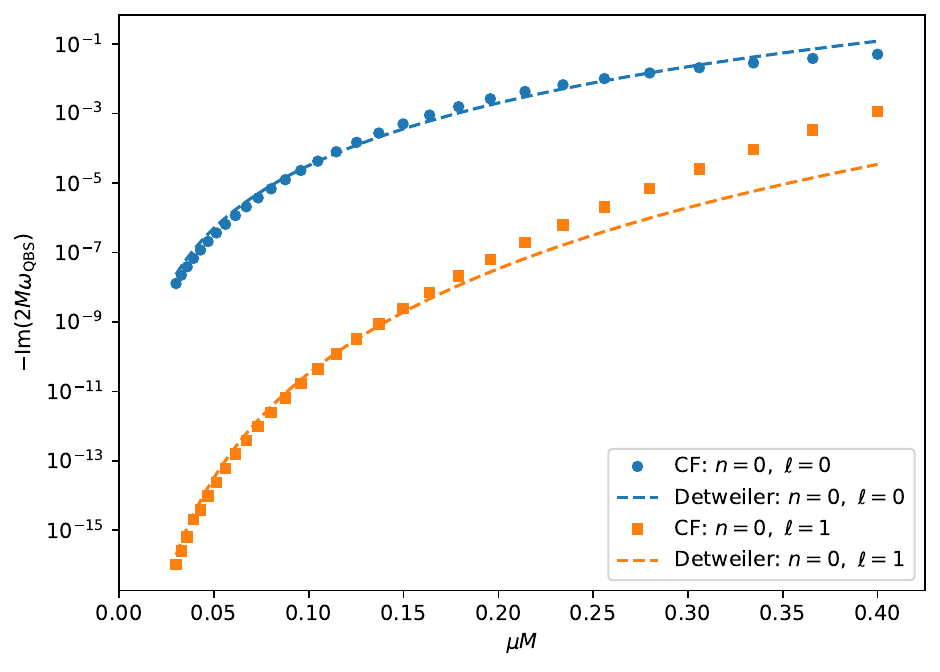}
      \caption{Damping magnitudes \(-\im(2M\omega_{\mathrm{QBS}})\) of the fundamental scalaron quasi-bound states, \(n=0\), for \(\ell=0,1\), as functions of the coupling \(\mu M\). Circles and squares show the continued-fraction results for \(\ell=0\) and \(\ell=1\), respectively; dashed curves show the corresponding Detweiler's hydrogenic approximations \cite{1980PhRvD..22.2323D}. The vertical axis is logarithmic. The numerical results approach the analytic curves at small coupling, \(\mu M\ll1\), while the differences increase as this approximation becomes less accurate at larger coupling.}
      \label{fig:qbs_mass_scan}
  \end{figure}

    For the quasi-bound states, we use the continued fraction equation consistent with the recurrence \ref{eq:scalar_branch_recurrence} with $\varepsilon=-1$. There are, of course, no known analytical solutions for these modes. However, in the small coupling (hydrogenic) regime, $\alpha=\mu M \ll 1$, Detweiler's matched asymptotic analysis gives a quasi-bound spectrum $\omega=\sigma+i\gamma$, with principal number $N=\ell+n+1$ \cite{1980PhRvD..22.2323D}. The real part follows from the hydrogenic
    binding energy (Eq.~(26) in \cite{1980PhRvD..22.2323D}),
    \begin{equation}
      \label{eq:detweiler_schw_sigma}
      \sigma \simeq \mu\left[1-\frac{\alpha^2}{2N^2}\right]\;.
    \end{equation}
    On the other hand, the imaginary part is controlled by the near-horizon absorption, with Detweiler's general Kerr expression (Eq. (28) in \cite{1980PhRvD..22.2323D}) reducing to the Schwarzschild limit by setting $a=0$, $r_+=2M$.
    The resulting expression simplifies considerably for the fundamental mode $n=0$. The case $\ell=0$ gives
    \begin{equation}
      -\im(2M\omega) = 32\alpha^6\;,
    \end{equation}
    while the $\ell=1$ case yields
    \begin{equation}
      -\im(2M\omega) = \frac{1}{3}\alpha^{10}\left(1+16\alpha^2\right) \;.
    \end{equation}
    It is these expressions that we have plotted in Fig~\ref{fig:qbs_mass_scan} for comparison with the root finding code. As $\mu M\to 0$ our root-finding results should asymptote to the hydrogenic curves above, providing a stringent consistency check of the implementation in the small-mass limit. As can be seen from Fig~\ref{fig:qbs_mass_scan}, agreement is generally good for very low $\mu M$, and degrades for larger $\mu M$ as the limit $\mu M\ll 1$ starts to break down.

We next validate the decaying-branch continued-fraction implementation against published Schwarzschild scalar quasi-bound-state frequencies, thereby providing a direct check against published values of the decaying-branch recurrence evaluated with its associated Nollert termination. As in the QNM comparison of Eq.~\eqref{eq:delta_cf_ref}, we quantify the agreement using the absolute complex-frequency difference
\begin{equation}
  \label{eq:delta_cf_ref_qbs}
\left|\Delta_{\rm CF-ref}^{\rm QBS}\right|
=
\left|
2M\omega_{\rm QBS}^{\rm this}
-
2M\omega_{\rm QBS}^{\rm ref}
\right|.
\end{equation}
        \begin{table}[t]
\centering
\small
\setlength{\tabcolsep}{4pt}
\renewcommand{\arraystretch}{1}

\begin{tabularx}{0.85\linewidth}{
|@{}
>{\centering\arraybackslash}p{0.06\linewidth}| %>{\centering\arraybackslash}p{0.13\linewidth}
>{\centering\arraybackslash}X|
>{\centering\arraybackslash}X|
>{\centering\arraybackslash}p{0.06\linewidth}|
>{\centering\arraybackslash}p{0.14\linewidth}
@{}|
}
  %\toprule
  \hline
%\((\ell,\mu M)\)
\(\ell\)
&
\(2M\omega_{\rm QBS}^{\rm this}\)
&
\(2M\omega_{\rm QBS}^{\rm ref}\)
&
Ref.
&
\(\Bigl|\Delta_{\rm CF-ref}^{\rm QBS}\Bigr|\)
\\
%\midrule
  \hline
\(0\)
&
\(\begin{gathered}
0.4954575282\\[-0.15em]
{}-0.0012730497\,i
\end{gathered}\)
&
\(\begin{gathered}
0.4954575\\[-0.15em]
{}-0.00127305\,i
\end{gathered}\)
&
\cite{Correa:2024xki}
&
\(2.821\times10^{-8}\)
\\ [15pt]
%&&&&\\
%\((1,0.25)\)
\(1\)
&
\(\begin{gathered}
0.4957823386\\[-0.15em]
{}-1.48725544\times10^{-6}\,i
\end{gathered}\)
&
\(\begin{gathered}
0.4957825\\[-0.15em]
{}-1.487255\times10^{-6}\,i
\end{gathered}\)
&
\cite{Correa:2024xki}
&
\(1.614\times10^{-7}\)
\\ [15pt]
%&&&&\\
%\((2,0.25)\)
\(2\)
&
\(\begin{gathered}
0.4982200441\\[-0.15em]
{}-3.91372745\times10^{-12}\,i
\end{gathered}\)
&
\(\begin{gathered}
0.4982200\\[-0.15em]
{}-4.509475\times10^{-12}\,i
\end{gathered}\)
&
\cite{Correa:2024xki}
&
\(4.408\times10^{-8}\)
\\
\hline
%\bottomrule
\end{tabularx}

\caption{Frequency-domain validation of selected scalar
Schwarzschild quasi-bound states at \(\mu M=0.25\). The reference
values are taken from Table~II of Ref.~\cite{Correa:2024xki}; the
conversion to the present \(2M\omega\) convention is detailed in
Appendix~\ref{app:frequencies_from_literature}. The differences are defined by Eq \eqref{eq:delta_cf_ref_qbs}.
The monopole entry \((\ell,\mu M)=(0,0.25)\) corresponds to the mode tabulated in Ref. ~\cite{Correa:2024xki}. The distinct fundamental monopole is reported in Table \ref{tab:td_qbs_validation} as part of the \(n=0\) comparison for \(\ell=0,1,2\).
}
\label{tab:qbs_frequency_domain_validation}
\end{table}
In Table~\ref{tab:qbs_frequency_domain_validation}, we show the comparison for $\ell=0,1,2$ at $\mu M=0.25$. The selected modes have damping magnitudes $-\operatorname{Im}(2M\omega_{\rm QBS})$ ranging from approximately $10^{-3}$ down to $10^{-12}$. For the $\ell=2$ mode, both results $2M\omega_{QBS}^{\rm this}$ and $2M\omega_{QBS}^{\rm ref}$ place the damping at order $10^{-12}$, although the associated imaginary parts differ at the coefficient level. Since the reference real part is available only to the precision of the printed value $0.99644$ in $\omega/\mu$, the absolute complex-frequency difference is controlled almost entirely by the rounding of that real part and thus does not provide a sensitive measure of the difference between the damping rates.

\subsection{Characteristic time-domain evolution and mode extraction}

We now turn from the frequency-domain eigenvalue problem of the previous section to the recovery of scalar QNM frequencies in the characteristic initial--worldtube evolution described in \S\ref{sec:time_domain_formulations_and_numerical_setup}. To this end, we first verify convergence of the underlying characteristic evolution. We then extract stable modal components using the matrix-pencil analysis and compare the resulting frequencies with the continued-fraction values reported above. We also examine their robustness under changes of the numerical setup and initial data.
Finally, we conclude the section with time-domain quasi-bound estimates and a demonstration of QNM and QBS recovery from a single finite-radius waveform.

The evolutions underlying the time-domain frequency benchmarks are performed without added Kreiss--Oliger dissipation. We consider non-zero dissipation coefficients separately in the robustness tests.
  
  \subsubsection{Convergence of the characteristic evolution}
  \label{sec:td_convergence}
\begin{figure}[!h]
    \centering
    \includegraphics[width=0.65\textwidth]{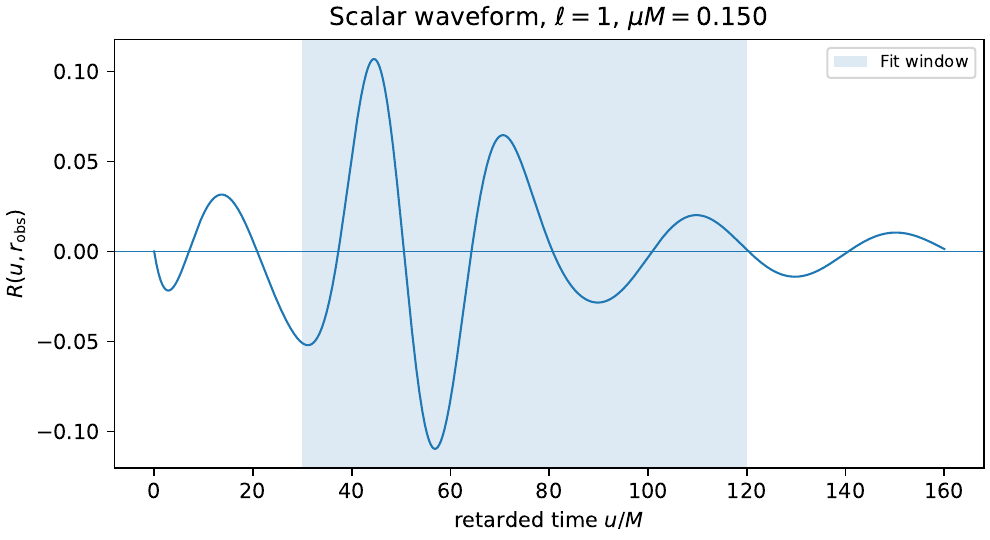}
    \caption{
      Representative finite-radius scalar waveform for $\ell=1$, $\mu M=0.15$ with $r_{max} = 1000M$ and $r_{obs} = 50M$. The shaded region, as a representative fit window, marks the interval ${30M\leq u\leq120M}$ is used for the global self-convergence comparison shown in Figure \ref{fig:td_scalar_convergence} and summarized in Table \ref{tab:td_convergence}.
    }
    \label{fig:td_waveform_convergence_window}
  \end{figure}

  Before performing the modal extraction, we verify convergence of the underlying time-domain solution. We use the representative scalar mode with $\ell=1$ and $\mu M=0.15$, which is also considered in the matrix-pencil and robustness analyses below. Fig \ref{fig:td_waveform_convergence_window} shows the corresponding finite radius waveform, with the shaded region highlighting the interval ($30M\leq u\leq120M$) used to evaluate the convergence diagnostics below. For this run, we set the outer boundary at \(r_{\max}=1000M\) and extract the waveform at \(r_{\rm obs}\simeq50M\). In this way, the estimated outer-boundary reflection time ($u_{refl}/M\simeq 1912$) lies far beyond the interval used in the convergence test.

  For the convergence test, we evolve the same physical setup at three nested resolutions, with a fixed refinement ratio of $2$ between successive grids. Let $R_h$, $R_{h/2}$, and $R_{h/4}$ denote the corresponding waveforms extracted at the same observation radius. We define
\begin{equation}
  Q =
  \frac{\|R_h-R_{h/2}\|_2}
       {\|R_{h/2}-R_{h/4}\|_2}\;,
  \qquad \text{and} \qquad
  p=\log_2 Q \;.
\end{equation}
Thus fourth-order convergence corresponds to \(Q\simeq16\) and
\(p\simeq4\).  The two convergence tests in
Table~\ref{tab:td_convergence} give \(p=4.012\) and \(p=4.001\),
respectively, consistent with the expected fourth-order behaviour of the
hypersurface solve and RK4 time integration.

\begin{figure}[!h]
      \centering
      \includegraphics[width=0.65\textwidth]{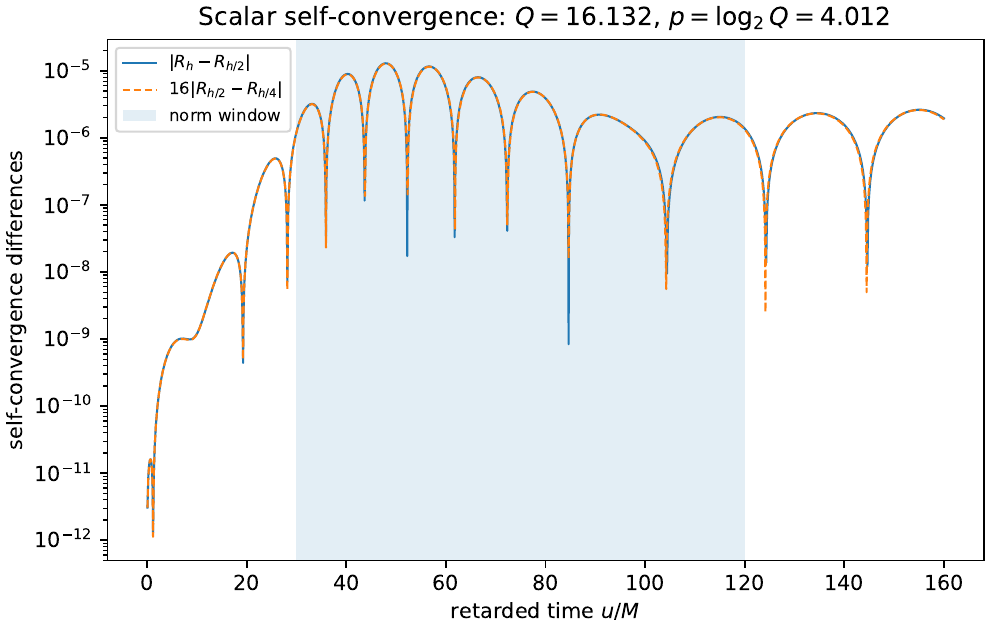}
      \caption{
Fourth-order self-convergence of the scalar time-domain evolution for
\(\ell=1\), \(\mu M=0.15\), with \(r_{\max}=1000M\) and
\(r_{\rm obs}\simeq50M\).  The coarse--medium difference
\(|R_h-R_{h/2}|\) is compared with the medium--fine difference
\(|R_{h/2}-R_{h/4}|\) rescaled by \(2^4=16\).  The agreement over
\({30M\le u\le120M}\) is consistent with the fourth-order convergence
reported in Table~\ref{tab:td_convergence}.
}
\label{fig:td_scalar_convergence}
    \end{figure}

Fig.~\ref{fig:td_scalar_convergence} shows the pointwise behaviour underlying the global self-convergence row of Table~\ref{tab:td_convergence}.  The figure shows the differences between successive resolutions, with the medium--fine difference rescaled by \(2^4=16\).  The overlap of the two curves over the selected representative ringdown interval provides a pointwise confirmation of
the fourth-order convergence summarized in
Table~\ref{tab:td_convergence}.

    \begin{table}[t]
\centering
\setlength{\tabcolsep}{3.5pt}
\renewcommand{\arraystretch}{1.18}

\begin{tabularx}{\linewidth}{|
>{\raggedright\arraybackslash}p{0.15\linewidth}|
>{\raggedright\arraybackslash}p{0.13\linewidth}|
Y|
>{\raggedright\arraybackslash}p{0.19\linewidth}|
>{\raggedright\arraybackslash}p{0.12\linewidth}|}
\hline
Diagnostic
& Case
& Parameters varied
& Fit window for $R(u,r_{\rm obs})$
& Outcome \\
\hline

\multirow[t]{2}{0.17\linewidth}{Self-convergence}
& Global
& \begin{tabular}[t]{@{}l@{}}
  $N_x=1001,2001,4001$\\
  $\Delta u/M=0.08,0.04,0.02$
  \end{tabular}
& $30M\le u\le120M$
& \begin{tabular}[t]{@{}l@{}}
  $Q=16.132$\\
  $p=4.012$
  \end{tabular}
\\
\cline{2-5}

& Time step
& \begin{tabular}[t]{@{}l@{}}
  fixed $N_x=3001$\\
  $\Delta u/M=0.08,0.04,0.02$
  \end{tabular}
& $30M\le u\le100M$
& \begin{tabular}[t]{@{}l@{}}
  $Q=16.011$\\
  $p=4.001$
  \end{tabular}
\\
\hline
\end{tabularx}

\caption{Convergence diagnostics for the scalar time-domain calculation.}
\label{tab:td_convergence}
\end{table}

        \subsubsection{QNM extraction and TD--CF comparison}

In \S\ref{sec:scalar_qnm_and_frequency_domain_validation} (Table~\ref{tab:scalar_cf_benchmarks}), we benchmarked the frequency-domain scalar spectrum against independent literature results. We now test whether the characteristic initial-value evolution independently recovers the same outgoing QNM frequencies in Table\ref{tab:scalar_cf_benchmarks}. In order to enable a direct row-by-row comparison, we apply the time-domain analysis to the same set of massless, low-mass, and finite-mass modes used in the frequency-domain benchmark table. 

\begin{figure}[t]
      \centering
      \includegraphics[width=\textwidth]{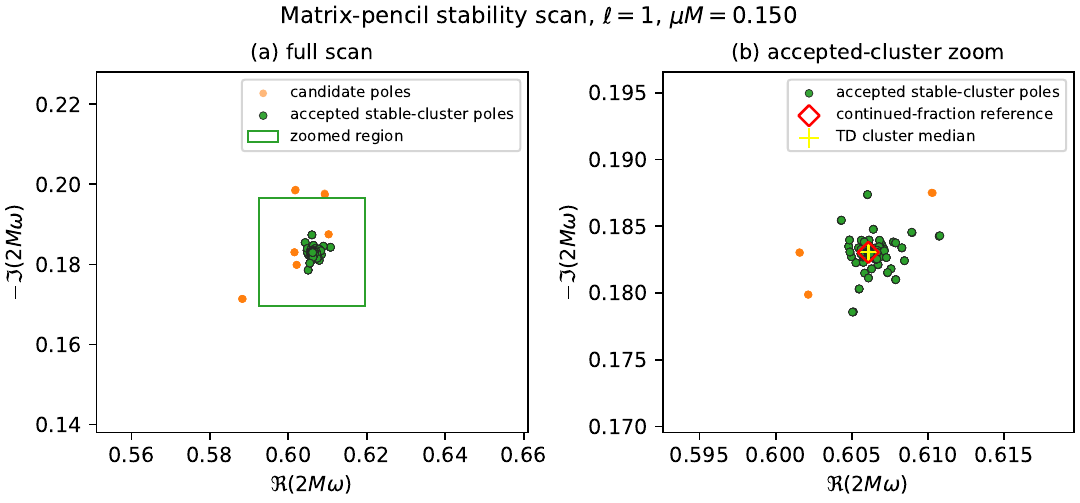}
      \caption{
Matrix-pencil stability scan for the scalar fundamental QNM with $\ell=1$ and $\mu M=0.15$. Panel (a) shows the full set of candidate poles obtained from the scan over fitting windows, pencil ranks, and decimation factors; the accepted stable-cluster poles are highlighted, and the rectangle indicates the region enlarged in panel (b). Panel (b) zooms into the accepted cluster and shows the cluster median, together with the independent continued-fraction QNM frequency. The horizontal and vertical coordinates are respectively $\Re(2M\omega)$ and $-\Im(2M\omega)$, so that damped modes appear at positive vertical coordinate. The continued-fraction result is shown only as an external reference and is not used in the matrix-pencil fit.
}
\label{fig:td_mode_stability}
    \end{figure}

Before presenting the table, we again consider the scalar fundamental mode with $\ell=1$ and $\mu M=0.15$ illustrated in Fig~\ref{fig:td_waveform_convergence_window}, to examine the poles obtained from the matrix-pencil stability scan, see Fig~\ref{fig:td_mode_stability}. Candidate poles recurring across the window, rank, and decimation scan form a compact accepted cluster in the complex-frequency plane. The accepted cluster lies close to the continued-fraction frequency of the outgoing scalar fundamental QNM.

The same time-domain-only selection procedure is applied to all modes
reported in Table~\ref{tab:scalar_td_benchmarks}. The table contains
the cluster median and componentwise median absolute deviations for
the representative case shown in
Fig.~\ref{fig:td_mode_stability}, together with the corresponding
results for the remaining modes in the matched benchmark set. The
massless block contains $\ell=0,\ldots,3$, the low-mass block contains
$\ell=0,\ldots,4$ at $\mu M=0.05$, and the finite-mass block contains
the $\ell=1,2$ modes at $\mu M=0.10$, $0.15$, and $0.20$.

For each mode, we quantify the agreement between the time-domain and
continued-fraction calculations using the absolute complex-frequency
difference
\begin{equation}
    \left|\Delta_{\mathrm{TD-CF}}\right|
    =
    \left|
        2M\omega_{\mathrm{TD}}^{\mathrm{this}}
        -
        2M\omega_{\mathrm{CF}}^{\mathrm{this}}
    \right|
    =
    \sqrt{
        \left(\Omega_{\mathrm{TD}}-\Omega_{\mathrm{CF}}\right)^2
        +
        \left(\Gamma_{\mathrm{TD}}-\Gamma_{\mathrm{CF}}\right)^2
    }.
    \label{eq:delta_td_cf}
\end{equation}
As before, the differences are evaluated using the unrounded frequencies rather
than the six-decimal values displayed in the table.

    \begin{table*}[t]
    \centering
    \begingroup
    \footnotesize
    \setlength{\tabcolsep}{3.5pt}
    \renewcommand{\arraystretch}{1.12}

    \begin{tabular}{|c|c|c|c|c|}
      \hline
        \toprule
        $(\ell,\mu M)$
        & $2M\omega_{\mathrm{TD}}^{\mathrm{this}}$
        & $2M\omega_{\mathrm{CF}}^{\mathrm{this}}$
        & $\left|\Delta_{\mathrm{TD-CF}}\right|$
        & $\left(\mathrm{MAD}_{\Omega},
                 \mathrm{MAD}_{\Gamma}\right)$
                 \\
                 \hline
        \midrule

        \multicolumn{5}{|c|}{
            \itshape Massless scalar modes
        }
        \\
        \hline
        \addlinespace[1pt]

        $(0,0.00)$
        & $0.221019-0.209852\,i$
        & $0.220910-0.209791\,i$
        & $1.24727\times10^{-4}$
        & $\left(4.39\times10^{-4},\,4.74\times10^{-4}\right)$
        \\

        $(1,0.00)$
        & $0.585872-0.195320\,i$
        & $0.585872-0.195320\,i$
        & $6.50922\times10^{-8}$
        & $\left(1.25\times10^{-7},\,1.22\times10^{-7}\right)$
        \\

        $(2,0.00)$
        & $0.967288-0.193517\,i$
        & $0.967288-0.193518\,i$
        & $4.24062\times10^{-7}$
        & $\left(1.07\times10^{-5},\,8.66\times10^{-6}\right)$
        \\

        $(3,0.00)$
        & $1.350731-0.192999\,i$
        & $1.350732-0.192999\,i$
        & $1.01188\times10^{-6}$
        & $\left(7.20\times10^{-6},\,8.31\times10^{-6}\right)$
        \\
        \hline
        \midrule

        \multicolumn{5}{|c|}{
            \itshape Low-mass scalar modes
        }
        \\
        \hline
        \addlinespace[1pt]

        $(0,0.05)$
        & $0.222811-0.204722\,i$
        & $0.221990-0.205692\,i$
        & $1.27067\times10^{-3}$
        & $\left(1.53\times10^{-3},\,1.32\times10^{-3}\right)$
        \\

        $(1,0.05)$
        & $0.588109-0.193976\,i$
        & $0.588109-0.193976\,i$
        & $3.25027\times10^{-7}$
        & $\left(1.21\times10^{-6},\,1.12\times10^{-6}\right)$
        \\

        $(2,0.05)$
        & $0.968866-0.192976\,i$
        & $0.968866-0.192976\,i$
        & $1.68299\times10^{-8}$
        & $\left(9.18\times10^{-6},\,7.27\times10^{-6}\right)$
        \\

        $(3,0.05)$
        & $1.351912-0.192714\,i$
        & $1.351912-0.192714\,i$
        & $1.64611\times10^{-7}$
        & $\left(1.18\times10^{-5},\,1.49\times10^{-5}\right)$
        \\

        $(4,0.05)$
        & $1.735766-0.192609\,i$
        & $1.735766-0.192609\,i$
        & $6.49393\times10^{-7}$
        & $\left(1.86\times10^{-5},\,2.30\times10^{-5}\right)$
        \\
        \hline
        \midrule

        \multicolumn{5}{|c|}{
            \itshape Finite-mass scalar modes
        }
        \\
        \hline
        \addlinespace[1pt]

        $(1,0.10)$
        & $0.594834-0.189911\,i$
        & $0.594831-0.189914\,i$
        & $4.03450\times10^{-6}$
        & $\left(2.40\times10^{-5},\,2.48\times10^{-5}\right)$
        \\

        $(2,0.10)$
        & $0.973607-0.191349\,i$
        & $0.973608-0.191349\,i$
        & $1.06170\times10^{-7}$
        & $\left(5.77\times10^{-5},\,4.67\times10^{-5}\right)$
        \\

        $(1,0.15)$
        & $0.606078-0.183055\,i$
        & $0.606080-0.183041\,i$
        & $1.41932\times10^{-5}$
        & $\left(1.70\times10^{-4},\,1.33\times10^{-4}\right)$
        \\

        $(2,0.15)$
        & $0.981512-0.188623\,i$
        & $0.981527-0.188624\,i$
        & $1.58377\times10^{-5}$
        & $\left(8.01\times10^{-5},\,6.70\times10^{-5}\right)$
        \\

        $(1,0.20)$
        & $0.621929-0.173006\,i$
        & $0.621914-0.173187\,i$
        & $1.81418\times10^{-4}$
        & $\left(7.17\times10^{-4},\,6.99\times10^{-4}\right)$
        \\

        $(2,0.20)$
        & $0.992595-0.184744\,i$
        & $0.992653-0.184778\,i$
        & $6.76150\times10^{-5}$
        & $\left(1.22\times10^{-4},\,1.49\times10^{-4}\right)$
        \\
        \bottomrule
        \hline
    \end{tabular}
    \endgroup
    \caption{
    Characteristic time-domain extractions of the fundamental scalar
    quasinormal frequency and comparison with the corresponding
    continued-fraction results. All frequencies are displayed in the
    convention
    $2M\omega=\Re(2M\omega)-i[-\Im(2M\omega)]$
    and are rounded to six decimal places. Here
    $\Omega=\Re(2M\omega)$ and
    $\Gamma=-\Im(2M\omega)>0$.
    The final column gives the componentwise cluster MADs. The differences
    $\left|\Delta_{\mathrm{TD-CF}}\right|$ are evaluated using the
    unrounded frequencies.
    }
    \label{tab:scalar_td_benchmarks}
\end{table*}

Across the listed massless, low-mass, and finite-mass cases, the time-domain cluster medians reproduce the continued-fraction frequencies within the stability scales of the accepted clusters. The largest TD--CF separations occur for the scalar monopole and for the more weakly damped finite-mass modes, where the accepted clusters are correspondingly broader. This is consistent with the increased difficulty of isolating a single exponentially damped component over a finite extraction interval.

\subsubsection{Robustness of the representative QNM extraction}

\label{sec:td_robustness}

We test the robustness of the characteristic extraction using the
representative \(\ell=1\), \(\mu M=0.15\) calculation employed in the
convergence and matrix-pencil analyses. Let
\(2M\omega_{\rm TD}^{\rm ref}\) denote the accepted-cluster median of
the reference calculation, and let
\[
W=[30M,120M]
\]
denote the common waveform-comparison interval. For each family of
variations, we define
\begin{equation}
\begin{aligned}
\Delta_{\omega}^{\max}
&=
\max_j
\left|
2M\omega_{\rm TD}^{(j)}
-
2M\omega_{\rm TD}^{\rm ref}
\right|,
\\
\delta_R^{\max}
&=
\max_j
\frac{
\left\|
R^{(j)}-R_{\rm ref}
\right\|_{2,W}
}{
\left\|
R_{\rm ref}
\right\|_{2,W}
}.
\end{aligned}
\label{eq:td_robustness_measures}
\end{equation}
Here \(j\) runs over the alternative parameter values in the
corresponding row of
Table~\ref{tab:td_robustness}. The waveform-level measure is reported
only when the two waveforms admit a meaningful direct comparison.

\begin{table}%[htbp]
\centering
\small
\setlength{\tabcolsep}{4pt}
\renewcommand{\arraystretch}{1.16}

\begin{tabularx}{\linewidth}{
|@{}
>{\raggedright\arraybackslash}p{0.18\linewidth}|
>{\raggedright\arraybackslash}X|
>{\centering\arraybackslash}p{0.15\linewidth}|
>{\centering\arraybackslash}p{0.15\linewidth}
@{}|
}
\hline%\toprule
Diagnostic
&
Parameters varied
&
\(\Delta_{\omega}^{\max}\)
&
\(\delta_R^{\max}\)
\\
\hline%\midrule

Inner worldtube
&
\(x_{\min}=-30,-35,-40\)
&
\(1.25\times10^{-6}\)
&
\(3.84\times10^{-3}\)
\\
\hline
Outer domain
&
\(r_{\max}/M=600,1000,1600\)
&
\(2.23\times10^{-6}\)
&
\(2.54\times10^{-2}\)
\\
\hline
Dissipation
&
KO strength \(=0,2.5\times10^{-3},5\times10^{-3}\)
&
\(3.82\times10^{-9}\)
&
\(5.05\times10^{-7}\)
\\
\hline
Extraction radius
&
\(r_{\rm obs}/M=40,50,60\)
&
\(2.93\times10^{-5}\)
&
--- 
\\
\hline
Initial data
&
\((r_0/M,\sigma/M)=(10,1),(12,1.5),(15,2)\)
&
\(1.72\times10^{-4}\)
&
---
\\

%\bottomrule
\hline
\end{tabularx}

\caption{Robustness diagnostics for the characteristic extraction of
the representative scalar fundamental QNM with
\(\ell=1\) and \(\mu M=0.15\). The reference calculation uses
\(x_{\min}=-35\), \(r_{\max}=1000M\), KO strength
\(2.5\times10^{-3}\), \(r_{\rm obs}=50M\), and
\((r_0/M,\sigma/M)=(12,1.5)\). The measures are defined in
Eq.~\eqref{eq:td_robustness_measures}; waveform differences are
evaluated over \(30M\leq u\leq120M\). No waveform-level difference is
reported for changes of extraction radius, which require additional
normalization and alignment, or for changes of initial data, which
deliberately modify the excitation amplitudes.}
\label{tab:td_robustness}
\end{table}

The accepted fundamental-mode cluster persists across all of the
variations in Table~\ref{tab:td_robustness}. Changes in the numerical
domain, dissipation strength, and extraction radius shift the cluster
median by at most \(2.93\times10^{-5}\). The largest overall shift,
\(1.72\times10^{-4}\), occurs under changes of the initial pulse and is
of the same order as the reference-cluster stability scales,
\[
\left(
\mathrm{MAD}_{\Omega},
\mathrm{MAD}_{\Gamma}
\right)
=
\left(
1.70\times10^{-4},
1.33\times10^{-4}
\right),
\]
reported in Table~\ref{tab:scalar_td_benchmarks}. This is consistent
with different initial data changing the relative excitation of QNM,
quasi-bound, prompt, and non-modal components without changing the
underlying eigenfrequencies.

The waveform as a whole is more sensitive to the finite numerical
domain than the extracted pole. In particular, varying \(r_{\max}\)
produces
\(\delta_R^{\max}=2.54\times10^{-2}\), while the corresponding
cluster-median shift is only
\(\Delta_{\omega}^{\max}=2.23\times10^{-6}\).
The full-waveform norm is more sensitive to the outer-domain variation than the extracted cluster median. Of course, this does not identify which waveform components account for the difference, but rather shows that the accepted QNM frequency remains stable despite the larger waveform-level variation.

\subsubsection{Time-domain quasi-bound extraction and simultaneous mode recovery}

We next examine the recovery of quasi-bound contributions from the characteristic evolution. Table~\ref{tab:td_qbs_validation} compares the time-domain estimates for \(\ell=0,1,2\) at \(\mu M=0.25\) with the corresponding \(n=0\) decaying-branch continued-fraction frequencies. For $\ell=0,1$, we report both frequency components $\re(2M\omega)$ and $\im(2M\omega)$. On the other hand, we only retain the real part $\re(2M\omega)$ for the $\ell=2$ case, whose damping is not resolved by the current extraction procedure. For each mode, we quantify the agreement between the time-domain and continued-fraction calculations using the absolute complex-frequency difference
\begin{equation}
  \left|\Delta_{\mathrm{TD-CF}}^{QBS}\right|
    =
    \left|
        2M\omega_{\mathrm{TD}}^{\mathrm{QBS}}
        -
        2M\omega_{\mathrm{CF}}^{\mathrm{QBS}}
    \right|
    =
    \sqrt{
        \left(\Omega_{\mathrm{TD}}-\Omega_{\mathrm{CF}}\right)^2
        +
        \left(\Gamma_{\mathrm{TD}}-\Gamma_{\mathrm{CF}}\right)^2
    }.
    \label{eq:delta_td_cf}
\end{equation}
with the differences evaluated using the unrounded frequencies rather
than the rounded values displayed in the table. Since no imaginary part $\im(2M\omega_{TD})$ is available for the $\ell=2$ case, we do not compute the associated difference $\left|\Delta_{\mathrm{TD-CF}}^{QBS}\right|$.

\begin{table}[t]
\centering
\small
\setlength{\tabcolsep}{5pt}
\renewcommand{\arraystretch}{1.15}

\begin{tabular}{|c|c|c|c|c|}
\hline %\toprule
\(\ell\)
& \(2M\omega_{\mathrm{CF}}\)
& \(2M\omega_{\mathrm{TD}}\)
& \(\mathrm{MAD}\)
& \(\Delta_{\mathrm{TD-CF}}^{\mathrm{QBS}}\) \\
\hline %\midrule

\multirow{2}{*}{\(0\)}
& \(0.48235028\)
& \(0.48231970\)
& \multirow{2}{*}{
    \(\bigl(1.17\times10^{-5},\,1.49\times10^{-5}\bigr)\)}
& \multirow{2}{*}{\(3.74\times10^{-5}\)}
\\
& \(-9.0706316\times10^{-3}\,i\)
& \(-9.0490349\times10^{-3}\,i\)
& &
\\
\hline %\addlinespace[5pt]

\multirow{2}{*}{\(1\)}
& \(0.49578233\)
& \(0.49578227\)
& \multirow{2}{*}{
    \(\bigl(6.22\times10^{-13},\,2.53\times10^{-13}\bigr)\)}
& \multirow{2}{*}{\(6.31\times10^{-8}\)}
\\
& \(-1.4872554\times10^{-6}\,i\)
& \(-1.4871359\times10^{-6}\,i\)
& &
\\
\hline %\addlinespace[5pt]

\multirow{2}{*}{\(2\)}
& \(0.4982200\)
& \(0.4982188\)
& \multirow{2}{*}{
    \(\bigl(1.64\times10^{-10},\,\text{\textemdash}\bigr)\)}
& \multirow{2}{*}{\textemdash}
\\
& \(-3.9137262\times10^{-12}\,i\)
& \textit{(real part only)}
& &
\\

\hline %\bottomrule
\end{tabular}

\caption{
Fundamental scalar quasi-bound frequencies at \(\mu M=0.25\)
(\(n=0\)), obtained from the decaying-branch continued fraction (CF)
and single-observer characteristic time-domain (TD) extractions.
The extraction radii are \(r_{\rm obs}/M\simeq28,64,144\) for
\(\ell=0,1,2\), respectively. TD entries are componentwise medians
of the selected matrix-pencil clusters. MAD pairs refer to
\(\Re(2M\omega)\) and \(-\Im(2M\omega)\), respectively, and measure
within-cluster spread rather than total numerical uncertainty.
For \(\ell=2\), only the TD real part and its MAD are reported.
We define \(\Delta_{\mathrm{TD-CF}}^{\mathrm{QBS}}
=|2M\omega_{\mathrm{TD}}-2M\omega_{\mathrm{CF}}|\), evaluated before
rounding; a dash indicates an unreported quantity. Displayed digits
specify central estimates, not guaranteed numerical accuracy.
}
\label{tab:td_qbs_validation}
\end{table}

    \begin{figure}[t]
      \centering
      \includegraphics[width=0.8\textwidth]{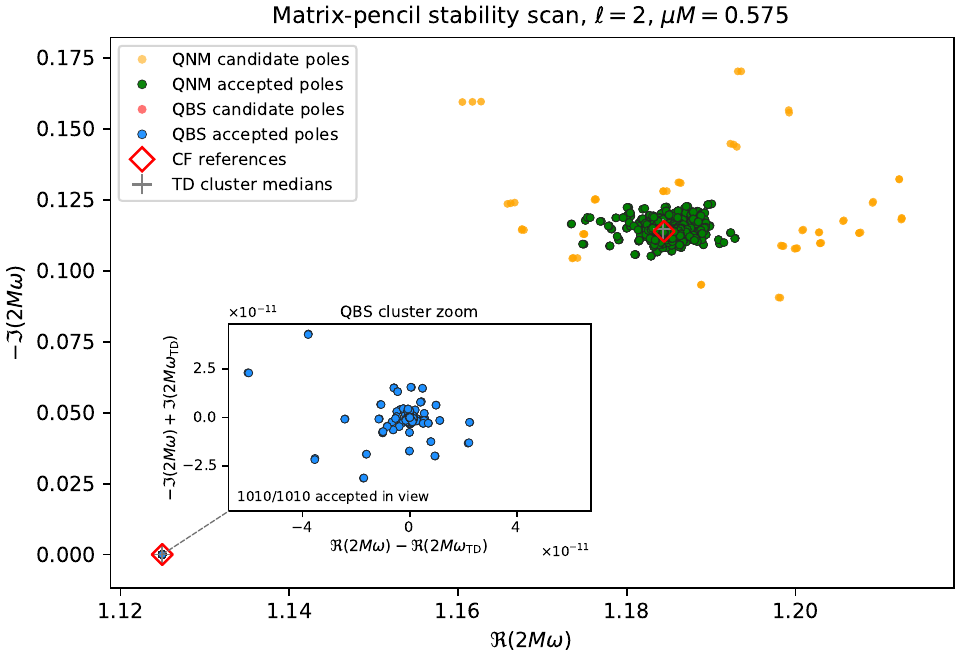}
      \caption{Recovery of the fundamental scalaron quasinormal mode (QNM) and quasi-bound state (QBS) from a single characteristic waveform, for \(\ell=2\), \(\mu M=0.575\), and extraction at \(r_{\mathrm{obs}}\simeq8M\). Separate matrix-pencil scans use early-time windows for the QNM and later, longer windows for the QBS. Candidate poles and accepted cluster members are distinguished as indicated in the legend; open diamonds denote continued-fraction (CF) frequencies and crosses denote the componentwise time-domain (TD) cluster medians. The main axes show \(\re(2M\omega)\) and \(-\im(2M\omega)\). The inset enlarges the accepted QBS cluster using offsets from its TD median and therefore displays within-cluster spread, not the TD–CF discrepancy. The absolute differences \(\lvert2M\omega_{\mathrm{TD}}-2M\omega_{\mathrm{CF}}\rvert\) are \(7.35\times10^{-4}\) for the QNM and \(4.46\times10^{-8}\) for the QBS. The CF values are introduced only after TD selection. Scan settings and numerical results are given in Table \ref{tab:simultaneous_extraction}.}
      \label{fig:simultaneous_mode_extraction_scan}
    \end{figure}
    Having demonstrated the time-domain extraction of QNMs and QBSs separately, we now turn to their simultaneous recovery from a single waveform using separate extraction windows. As mentioned above (\S\ref{sec:mode_content_identification}), within the time-domain characteristic evolution, no independent outer boundary is imposed at $r_{max}$, so that $r_{max}$ acts as a finite-domain truncation rather than as a spectral mode-selection boundary. Consequently, the resulting waveform may contain signatures of both QNMs and QBSs, with their relative prominence depending on the choice of initial data and extraction radius. This makes it possible, in principle, to recover both QNM and QBS frequencies from a single waveform. In practice, however, robust recovery of both complex frequencies from a single waveform is not equally straightforward throughout the $(\ell, \mu M)$ parameter space. For example, a very long-lived QBS may have a well-resolved oscillation frequency while its damping rate remains poorly constrained, since many oscillation cycles can occur over an interval in which the amplitude changes only weakly \cite{PhysRevLett.109.081102}. On the other hand, a QNM may be harder to isolate when appreciable QBS contributions obscure its damped oscillations, thus hindering the determination of its oscillation frequency and damping rate \cite{Decanini:2015yba}. For the illustrative case below, we select \(\ell=2\) and \(\mu M=0.575\). With the chosen initial data and extraction radius, the QNM remains identifiable at early times, while the later QBS signal exhibits sufficient decay for its damping rate to be determined.

Fig~\ref{fig:simultaneous_mode_extraction_scan} presents the results of the simultaneous extraction. Both the QNM and the fundamental QBS are recovered from the same waveform at \(r_{\rm obs}\simeq8M\), using the fitting windows and scan settings listed in Table~\ref{tab:simultaneous_extraction}. We apply the same mode-identification procedure as in the preceding separate analyses, targeting the QNM in early-time windows and the QBS in later, longer windows. The extracted QNM and QBS frequencies differ from their continued-fraction benchmarks by \(7.35\times10^{-4}\) and \(4.46\times10^{-8}\), respectively, in \(2M\omega\). The corresponding relative differences in the damping rates are approximately \(0.63\%\) and \(1.2\times10^{-3}\%\). Thus, for the selected case, both the oscillation frequency and damping rate of each mode are recovered from a single characteristic waveform at a single extraction radius.

\begin{table}%[htbp]
  \centering
  \begin{tabular}{|l|c|c|}
\hline
Setting / diagnostic & QNM & QBS \\
\hline

\multirow{3}{*}{Window starts, \(u_{\rm start}/M\)}
& \(0,1,2,3,4,5,6,8,\)
& \(2000,2750,3500,4250,5000,\)
\\

& \(10,12,15,18,20,24,28,32\)
& \(5750,6500,7250,8000,8750,\)
\\
&
& \(9500,10250,11000\)
\\
\hline %\addlinespace

\multirow{2}{*}{Window lengths, \(L/M\)}
& \(15,20,25,30,35,\)
& \(7000,7750,8500,9250,10000,\)
\\
& \(40,45,50,60\)
& \(10750,11500,12250,13000\)
\\
\hline %\addlinespace

Matrix-pencil ranks
& \(4,6,8,10,12,14,16\)
& \(8,10,12,14,16,18,20\) \\
\hline

Decimation factors
& \(1,2,3\)
& \(16,18\) \\
\hline

Accepted cluster members
& \(4078\)
& \(1010\)
\\
\hline %\addlinespace

Distinct fitting combinations
& \(2039\)
& \(505\)
\\
\hline %\addlinespace

Componentwise MADs
& \(\bigl(7.9\times10^{-4},\,7.5\times10^{-4}\bigr)\)
& \(\bigl(3.7\times10^{-14},\,4.1\times10^{-14}\bigr)\)
\\
\hline %\addlinespace

\(2M\omega_{\mathrm{TD}}\)
& \(1.18426822-0.11465371\,i\)
& \(1.12493278-5.41685764\times10^{-5}\,i\)
\\
\hline%addlinespace

\(2M\omega_{\mathrm{CF}}\)
& \(1.18443122-0.11393729\,i\)
& \(1.12493282-5.41692409\times10^{-5}\,i\)
\\
\hline%addlinespace

\(\left|2M\omega_{\mathrm{TD}}
       -2M\omega_{\mathrm{CF}}\right|\)
& \(7.35\times10^{-4}\)
& \(4.46\times10^{-8}\)
\\
\hline %\addlinespace
\end{tabular}
\caption{
Fitting settings and cluster diagnostics for the simultaneous QNM
and QBS extractions.  Both use the same finite-radius waveform from
one characteristic evolution at \(r_{\rm obs}\simeq8M\).
The componentwise MADs are quoted in the \(2M\omega\) convention. As before, the TD--CF differences are evaluated using the unrounded frequencies.
}
\label{tab:simultaneous_extraction}
\end{table}

%%%%%%%%%%%%%%%%%%%%%%%%%%%%%%%%%%%%%%%%%%%%%%%%%%%%%%%%%%%%%%%%%%%%%%%%%%%%%%%%%%%

\section{Concluding remarks}
\label{sec:concluding_remarks}

In this work, we have studied scalaron quasinormal and quasi-bound modes of Schwarzschild black holes in metric \(f(R)\) gravity directly within the Bondi–Sachs formulation, using complementary frequency-domain and characteristic time-domain approaches. Using the same elimination procedure outlined in \cite{Mongwane:2024vao}, we have shown that the tensor master equation reduces to its general relativistic form, with the intrinsic tensor QNM spectrum thus remaining that of Schwarzschild in GR. The linearised Bondi-Sachs equations lead naturally to a closed master equation for the scalaron. The form of this equation shows that, for the viable analytic $f(R)$ models considered here, the scalar spectra depend on the theory only through the effective mass $\mu$. It then follows that, at fixed geometric mass \(M\), models with the same scalaron mass $\mu$ have identical linear scalar QNM and quasi-bound spectra. This degeneracy is restricted to the homogeneous linear spectral problem and does not, in general, extend to sourced or nonlinear dynamics. 

In the frequency-domain, we formulated the outgoing-QNM and decaying-QBS conditions within a common Frobenius, and continued-fraction framework, with explicit asymptotic branch prescriptions together with the corresponding Nollert remainders. These boundary conditions are equivalently encoded in branch-dependent Wronskian zeros, establishing a direct connection between the continued-fraction roots and the poles of the frequency-domain resolvent. We compared our frequencies with published QNM and QBS results thus providing quantitative validation of the methods adopted here. We also find that the quasi-bound damping rates approach Detweiler's approximation at small coupling, recovering a smooth connection between the continued-fraction solutions and the known analytic limit.

In the time-domain, the characteristic initial--worldtube evolution provides an independent test of the frequency-domain spectral description. Unlike the frequency-domain two-point boundary-value problems, the initial--worldtube evolution does not impose an outgoing or decaying spectral branch at the outer edge. The modal contributions are instead identified from their persistence across matrix-pencil scans, with the continued-fraction frequencies introduced only after the time-domain selection. The representative convergence tests recover the expected fourth-order behaviour, while the fundamental QNM frequencies extracted across the benchmark set agree with the continued-fraction values within the extraction-stability scales of the accepted clusters. Moreover, for the representative finite-mass calculation, the extracted pole also remains robust under variations of the numerical setup and initial data. We also use the time-domain analysis to recover quasi-bound contributions. The QBS results illustrate the limits of the current extraction procedure. At \(\mu M=0.25\), both the oscillation frequencies and damping rates are recovered for \(\ell=0,1\), whereas only the oscillation frequency is constrained for \(\ell=2\) while the damping rate remains unresolved. A recoverable quasi-bound oscillation therefore does not necessarily imply a reliable measurement of its decay rate. More generally, the reported cluster MADs measure sensitivity to the extraction choices, not total numerical uncertainty, and should be interpreted accordingly. Similarly, the robustness tests concern numerical pole recovery for a fixed evolution equation; they do not constitute a pseudospectral analysis of perturbations to the underlying operator. 

Using the representative case \(\ell=2\), \(\mu M=0.575\), in the time-domain, we have also demonstrated that it is feasible to extract both the fundamental QNM and QBS frequencies from the same waveform generated by a single initial--worldtube evolution. The extraction procedure in this case employs early-time fitting windows to capture the QNM response, and later, longer windows for the quasi-bound contribution. In this way, the two spectral families are obtained without explicitly enforcing the spectral condition associated with either branch at the finite outer boundary. This example demonstrates that the characteristic evolution and extraction procedure can resolve both the relatively rapid QNM damping and the much slower quasi-bound decay within a single waveform.

The results obtained here indicate that the characteristic formulation provides a versatile framework for perturbative studies.  It would be worthwhile to further expand on this work to rotating Kerr black holes. Other interesting avenues to pursue, relevant to $f(R)$ gravity is wave-form extraction. The finite-radius scalaron diagnostics employed here provide a starting point for a broader waveform-extraction framework that is adapted to massive fields. Beyond linear order in perturbation theory, a further question is whether long-lived scalaron configurations can source a delayed tensor response through scalaron–tensor coupling. Higher-order perturbations (and backgrounds with non-constant Ricci curvature) could also probe derivatives of \(f(R)\) that are absent from the current eigenvalue problem, thereby breaking the spectral degeneracy that exists at linear order.

\section{Acknowledgements}

SZ acknowledges funding from the National Astrophysics and Space Science Programme. DFM thanks the Research Council of Norway for their support and the resources provided by UNINETT Sigma2 -- the National Infrastructure for High-Performance Computing and Data Storage in Norway.

\appendix

\section{Tensor Nollert remainder and numerical consistency check}
\label{app:tensor_validation}

The tensor Frobenius ansatz, its three-term recurrence and the large order behaviour of the coefficient ratio are derived in \cite{Mongwane:2024vao}. In the following we summarize only the recurrence equations
needed to specify the Nollert termination adopted in the current implementation, and to present its numerical validation. The recurrence is
\begin{subequations}
\label{eq:tensor_recurrence_qnm}
\begin{align}
\alpha_0a_1+\beta_0a_0 &=0,\\
\alpha_ka_{k+1}+\beta_ka_k+\gamma_ka_{k-1}&=0,
\qquad k\geq1,
\end{align}
\end{subequations}
with
\begin{subequations}
\label{eq:tensor_recurrence_coefficients}
\begin{align}
  \alpha_k
  &=
  -k^2+(4M\rho-4)k+4M\rho-3,\\
  \beta_k
  &=
  2k^2-(16M\rho-2)k
  +32M^2\rho^2-8M\rho-3+\ell(\ell+1),\\
  \gamma_k
  &=
  -k^2+(8M\rho+2)k
  -16M^2\rho^2-8M\rho.
\end{align}
\end{subequations}
Defining the coefficient ratio
\begin{equation}
  R_k\equiv-\frac{a_{k+1}}{a_k},
  \label{eq:tensor_ratio_definition}
\end{equation}
Eq.~\eqref{eq:tensor_recurrence_qnm} gives
\begin{equation}
  R_{k-1}
  =
  \frac{\gamma_k}{\beta_k-\alpha_kR_k}.
  \label{eq:tensor_ratio_recurrence}
\end{equation}
At a large truncation index \(N\), we seek an asymptotic remainder of
the form
\begin{equation}
  R_N
  \sim
  C_0+\frac{C_1}{\sqrt{N}}+\frac{C_2}{N}
  +\mathcal{O}\!\left(N^{-3/2}\right).
  \label{eq:RN_asymptotic_general}
\end{equation}
with the coefficients $C_k$ determined by the large–$k$ behaviour of $\alpha_k$, $\beta_k$ and $\gamma_k$.  For the recurrence \eqref{eq:tensor_recurrence_qnm}, it follows that
\begin{equation}
  C_0 = -1,\qquad
  C_1^2 = -4M\rho,\qquad
  C_2 = 4M\rho + \frac{11}{4}\,.
\end{equation}
with \(\Re(C_1)>0\), which selects the minimal solution of the large-\(k\) recurrence. This branch is continued smoothly along the tracked overtone sequence. The remainder can then be written explicitly as
\begin{equation}
\label{eq:tail_tensor} 
  R_N \simeq -1
  + \frac{2\sqrt{-M\rho}}{\sqrt{N}}
  + \frac{4M\rho + 11/4}{N}\,.
\end{equation}
Numerically, one can either use \eqref{eq:tail_tensor} at $k=N$ as the starting value for a backward iteration of the ratio recurrence $R_{k-1} = \gamma_k / (\beta_k - \alpha_k R_k)$ down to the desired $k$, or
equivalently, replace the last denominator in the truncated continued fraction by $\beta_N - \alpha_N R_N$.  This Nollert-type termination accelerates and stabilizes the evaluation of highly damped overtones, and is now standard in high–precision calculations of quasinormal spectra (see, e.g., Refs.~\cite{1993PhRvD..47.5253N,Berti:2009kk,Konoplya:2011qq}).

Since the tensor recurrence is used here only as a numerical consistency check, and given that the Nollert remainder retained in Eq.~\eqref{eq:tail_tensor} is independent of \(\ell\) through \(\mathcal{O}(N^{-1})\), in what follows, we restrict the numerical validation to the representative multipole \(\ell=2\).

For each mode in Table~\ref{tab:tensor_high_overtone_validation}, we evaluated the continued fraction at $N_{\rm cut}=8000$, $16000$ and  $32000$, with the reported values being obtained with \(N_{\rm cut}=32000\). We define the difference from the literature value $\Delta_{\rm CF-ref}$ in the same way as Eq~\eqref{eq:delta_cf_ref}. We further define the truncation-stability measure of the obtained CF values by
\begin{equation}
\delta_N
=
\left|
2M\omega_{\rm CF}^{(32000)}
-
2M\omega_{\rm CF}^{(16000)}
\right|.
\label{eq:tensor_validation_measures}
\end{equation}
In each case, we verified that the change between \(N_{\rm cut}=16000\) and \(32000\) is smaller than that between \(N_{\rm cut}=8000\) and \(16000\).
\begin{table}[t]
\centering
%\scriptsize
\setlength{\tabcolsep}{3pt}
    \begin{tabular}{|@{}c|c|c|c|c@{}|}
  \hline
\(n\)
&
\(2M\omega_{\rm CF}^{\rm this}\)
&
\(2M\omega_{\rm ref}\)
&
\(\left|\Delta_{\rm CF-ref}\right|\)
&
\(\delta_N\)
\\
  \hline

\multicolumn{5}{|c|}{
\emph{Recent high-precision benchmark:
Ref.~\cite{Hirano:2024fgp}}
}
\\
\hline

\(10\)
&
\(\begin{gathered}
0.153106925771975\\[-0.15em]
{}-5.121653234763009\,i
\end{gathered}\)
&
\(\begin{gathered}
0.153106925771972\\[-0.15em]
{}-5.121653234763011\,i
\end{gathered}\)
&
\(3.44874\times10^{-15}\)
&
\(1.37346\times10^{-12}\)
\\
\hline

\multicolumn{5}{|c|}{
\emph{Published high-overtone comparison:
Ref.~\cite{1985RSPSA.402..285L}}
}
\\

\hline
\(29\)
&
\(\begin{gathered}
0.166108188308599\\[-0.15em]
{}-14.676270527604345\,i
\end{gathered}\)
&
\(\begin{gathered}
0.165814\\[-0.15em]
{}-14.677118\,i
\end{gathered}\)
&
\(8.97082\times10^{-4}\)
&
\(2.31528\times10^{-11}\)
\\

\(59\)
&
\(\begin{gathered}
0.146794671555976\\[-0.15em]
{}-29.696064765072727\,i
\end{gathered}\)
&
\(\begin{gathered}
0.148484\\[-0.15em]
{}-29.696417\,i
\end{gathered}\)
&
\(1.72566\times10^{-3}\)
&
\(1.33589\times10^{-8}\)
\\
%\bottomrule
  \hline
\end{tabular}
\caption{Selected \(\ell=2\) tensor Schwarzschild QNMs used to check
the continued-fraction solver and its Nollert termination. The
\(n=10\) reference frequency is taken from
Ref.~\cite{Hirano:2024fgp}; the value reported there as \(M\omega\)
has been multiplied by two to match the current \(2M\omega\)
normalisation. The \(n=29\) and \(n=59\) reference frequencies are
the values printed in Table~1 of Ref.~\cite{1985RSPSA.402..285L}. Leaver
uses the one-based convention \(n_{\rm L}=1\) for the fundamental,
so that \(n=n_{\rm L}-1\) for the \(\ell=2\) sequence. The quantities
\(\left|\Delta_{\rm CF-ref}\right|\) and \(\delta_N\), defined in Eqs \eqref{eq:delta_cf_ref} and \eqref{eq:tensor_validation_measures} respectively, are evaluated
using the unrounded roots.}
\label{tab:tensor_high_overtone_validation}
\end{table}
The obtained \(n=10\) result differs with the recent high-precision value of \cite{Hirano:2024fgp} by \(3.44874\times10^{-15}\), while its conservative truncation-stability
measure is \(1.37346\times10^{-12}\). At the same time, the \(n=29\) and \(n=59\) roots are internally well converged under increasing \(N_{\rm cut}\) as evidenced by their $\delta_N$ differences, but their $\Delta_{CF-ref}$ differences from Leaver's values are larger than the rounding scale of the published table. We therefore regard these two rows as deep-overtone checkpoints rather than as precision benchmarks.

Lastly, as an additional high-overtone stress test, we continued the same \(\ell=2\) sequence through \(n=100\). The \(N_{\rm cut}=32000\) result, which remained unchanged at $N_{\rm cut}=64000$ within the precision of the numerical output, is
\begin{equation}
2M\omega_{2,100}
=
0.134219374229091
-
50.207343362786160\,i.
\label{eq:tensor_l2_n100}
\end{equation}
The corresponding truncation difference for this value is $\delta_N = 3.48394\times10^{-7}$. The \(n=100\) result is included only as an endpoint consistency check of the Nollert-terminated backward
recurrence.

Together, Table \ref{tab:tensor_high_overtone_validation} and the truncation stability of the \(n=100\) endpoint provide a compact validation of the tensor continued-fraction implementation.
We only use these numerical tensor results for this consistency check; the physical analysis in the main text concerns the scalar QNM, quasi-bound, and characteristic time-domain sectors.

\section{Characteristic radial discretization}
\label{app:characteristic_discretization}

In this appendix we record the implementation details of the radial
hypersurface solve used in the scalar characteristic evolutions. The
continuum first-order system is given in
Sec.~\ref{sec:characteristic_evolution_scheme}. At each retarded-time step it
has the hierarchical form
\begin{equation}
R_{,u}=\psi,
\label{eq:app_td_evolution}
\end{equation}
together with the radial hypersurface equation
\begin{equation}
\psi_{,r}
+
\frac{\psi}{r}
=
\frac{r-2M}{2r}R_{,rr}
+
\frac{r-M}{r^2}R_{,r}
-
\frac{\ell(\ell+1)+\mu^2r^2}{2r^2}R.
\label{eq:app_td_hypersurface}
\end{equation}

\subsection{Logarithmic radial coordinate}

To place the inner worldtube close to the future horizon while
retaining a uniform numerical grid, we use
\begin{equation}
x
=
\log\left(\frac{r}{2M}-1\right),
\qquad
q\equiv r-2M=2Me^x,
\qquad
r=2M+q,
\qquad
\frac{dr}{dx}=q.
\label{eq:app_log_coordinate}
\end{equation}
The radial grid is therefore
\begin{equation}
x_i=x_{\min}+i\Delta x,
\qquad
i=0,\ldots,N_x-1,
\label{eq:app_x_grid}
\end{equation}
with
\begin{equation}
q_i=2Me^{x_i},
\qquad
r_i=2M+q_i.
\end{equation}

The required radial derivatives may be written as
\begin{equation}
R_{,r}
=
\frac{1}{q}R_{,x},
\qquad
R_{,rr}
=
\frac{R_{,xx}-R_{,x}}{q^2}.
\label{eq:app_derivative_transform}
\end{equation}
Defining
\begin{equation}
y\equiv r\psi,
\label{eq:app_y_definition}
\end{equation}
the left-hand side of
Eq.~\eqref{eq:app_td_hypersurface} becomes
\[
(r\psi)_{,r}=y_{,r}.
\]
Multiplication by \(q=dr/dx\) then reduces the hypersurface equation
to
\begin{equation}
y_{,x}=F_R,
\label{eq:app_integrated_hypersurface}
\end{equation}
where
\begin{equation}
F_R
=
\frac{1}{2}R_{,xx}
+
\frac{q}{2r}R_{,x}
-
\frac{q}{2r}
\left[
\ell(\ell+1)+\mu^2r^2
\right]R.
\label{eq:app_scalar_source}
\end{equation}
This form is regular as \(x\rightarrow-\infty\), since it introduces
no division by \(r-2M\).

\subsection{Fourth-order cumulative quadrature}

At each Runge--Kutta stage, the stage value of \(R\) is known at all
radial grid points. We evaluate the derivatives in
Eq.~\eqref{eq:app_scalar_source} using fourth-order finite-difference
operators,
\begin{equation}
R_{,x}(x_i)
=
(D_xR)_i+\mathcal{O}(\Delta x^4),
\qquad
R_{,xx}(x_i)
=
(D_{xx}R)_i+\mathcal{O}(\Delta x^4),
\label{eq:app_fd_operators}
\end{equation}
with centred stencils in the interior and one-sided fourth-order
closures near the boundaries. These derivatives determine the
discrete source
\begin{equation}
(F_R)_i
=
\frac{1}{2}(D_{xx}R)_i
+
\frac{q_i}{2r_i}(D_xR)_i
-
\frac{q_i}{2r_i}
\left[
\ell(\ell+1)+\mu^2r_i^2
\right]R_i.
\label{eq:app_discrete_source}
\end{equation}

Integrating Eq.~\eqref{eq:app_integrated_hypersurface} from
\(x_0=x_{\min}\) to \(x_i\) gives
\begin{equation}
y_i
=
y_0
+
\int_{x_0}^{x_i}F_R(x)\,dx.
\label{eq:app_y_integral}
\end{equation}
We approximate the integral using a fourth-order
Euler--Maclaurin-corrected composite trapezoidal rule. Let
\begin{equation}
T_i
=
\Delta x
\left[
\frac{1}{2}(F_R)_0
+
\sum_{j=1}^{i-1}(F_R)_j
+
\frac{1}{2}(F_R)_i
\right]
\label{eq:app_cumulative_trapezoid}
\end{equation}
denote the cumulative trapezoidal approximation. The leading
Euler--Maclaurin correction gives
\begin{equation}
I_i
=
T_i
-
\frac{\Delta x^2}{12}
\left[
(D_xF_R)_i-(D_xF_R)_0
\right],
\label{eq:app_euler_maclaurin}
\end{equation}
where \(D_xF_R\) is evaluated using the same fourth-order derivative
operators. For a sufficiently smooth source,
\begin{equation}
I_i
=
\int_{x_0}^{x_i}F_R(x)\,dx
+
\mathcal{O}(\Delta x^4).
\end{equation}

The hypersurface variable is then reconstructed by
\begin{equation}
y_i=y_0+I_i,
\qquad
\psi_i=\frac{y_i}{r_i}.
\label{eq:app_hypersurface_reconstruction}
\end{equation}
The inner value \(y_0=r_{min}\psi_0\) is supplied by the inner-worldtube
condition in
Eq.~\eqref{eq:td_inner_bc}. It is reevaluated from
the corresponding stage value of \(R\) at every Runge--Kutta
substep. No independent outer-boundary value of \(\psi\) is required:
Eq.~\eqref{eq:app_hypersurface_reconstruction} determines it
throughout the radial grid by outward integration from the
worldtube.

\subsection{Runge--Kutta update}

The resulting semi-discrete evolution equation is
\begin{equation}
\frac{dR_i}{du}
=
\psi_i,
\label{eq:app_semidiscrete_evolution}
\end{equation}
with any Kreiss--Oliger dissipation added to the right-hand side of
this evolution equation rather than to the radial hypersurface
solve. The classical explicit fourth-order Runge--Kutta method is
used for the integration in \(u\).

At each Runge--Kutta stage, the numerical operations are therefore:

\begin{enumerate}
\item evaluate \(R_{,x}\) and \(R_{,xx}\) from the current stage
      values of \(R_i\);
\item construct the source \((F_R)_i\) from
      Eq.~\eqref{eq:app_discrete_source};
\item evaluate \(D_xF_R\) and the cumulative corrected integral
      \(I_i\);
\item reconstruct \(y_i\) and \(\psi_i\) from
      Eq.~\eqref{eq:app_hypersurface_reconstruction};
\item use \(\psi_i\), together with any chosen dissipation term, as
      the Runge--Kutta right-hand side for \(R_i\).
\end{enumerate}

The radial finite-difference operators, corrected cumulative
quadrature, and retarded-time integration are all fourth-order
accurate. The resulting global convergence order is verified
numerically in Sec.~\ref{sec:td_convergence}.

  \section{Literature benchmarks and frequency conventions}
  \label{app:frequencies_from_literature}

  The reference conventions and conversions are as follows:
\begin{itemize}
    \setlength{\itemsep}{3pt}

    \item
    \textbf{Cavalcante and da Cunha \cite{Cavalcante:2021scq}.}
    The massless entries are the fundamental scalar Schwarzschild
    values obtained in the $q=Q=0$ limit using the isomonodromic
    $\tau_V$-function method. Their frequencies are quoted as
    $M\omega$ and have therefore been multiplied by two to match the
    present $2M\omega$ convention. The $\ell=0$ result is taken from
    their Table II, while the $\ell=1,2,3$ results are taken from
    their Table I. These four modes form the massless normalization
    block used in both Tables~\ref{tab:scalar_cf_benchmarks} and
    \ref{tab:scalar_td_benchmarks}.

    \item
    \textbf{Alves et al \cite{Alves:2025erh}.}
    The low-mass entries are the fundamental ($n=0$), $m=0.1$
    values reported in their Table III, converted to the present
    normalization as described above. In our notation this
    corresponds to $\mu M=0.05$. We retain the multipoles
    $\ell=0,\ldots,4$, matching the low-mass modes reported in the
    characteristic time-domain comparison. The $\ell=5,6$ entries
    are not included because they are quoted at lower source
    precision and are not part of the retained time-domain
    benchmark set.

    The comparatively larger difference for
    $(\ell,\mu M)=(0,0.05)$ in Table \ref{tab:scalar_cf_benchmarks} is associated with the finite-order
    Hill-determinant value tabulated in Ref.~\cite{Alves:2025erh}.
    The publicly archived Leaver implementation accompanying that work gives \cite{alves_2025_17580446}
    \[
    2M\omega
    = 0.2219897934 - 0.2056931072\,i
    \]
    at truncation order $N=100$, reducing the difference with the present
    continued-fraction result to $1.48\times10^{-6}$.  Increasing the
    truncation order in the same recurrence converges to the value obtained
    here.  We therefore do not interpret the $10^{-5}$ difference from the
    tabulated Hill value as a discrepancy between this work and \cite{Alves:2025erh}.

    \item
    \textbf{Dolan \cite{Dolan:2007mj}.}
    The $\ell=1$ entries at $\mu M=0.10$ and $0.20$ are the
    fundamental Schwarzschild results obtained from the $a=0$ rows
    of Appendix B, Table II. Although the corresponding source table carries
    an azimuthal-mode label, the spectrum is independent of that
    label in the Schwarzschild limit. Dolan quotes the frequencies
    as $M\omega$, so the reported values have been multiplied by
    two. Together with the $\mu M=0.15$ result described below,
    these entries provide a matched $\ell=1$ sequence over the
    three finite masses used in the time-domain study.

    \item
    \textbf{Konoplya, Matyjasek and Zhidenko \cite{Konoplya:2026rjh}.}
    The $\ell=1$, $\mu M=0.15$ entry in Table I is their accurate Frobenius
    result at $\mu r_0=0.30$, where $r_0=2M$ is the Schwarzschild
    horizon radius. Consequently, their mass parameter corresponds
    to $\mu M=0.15$, and their tabulated frequency
    $r_0\omega$ is already in the $2M\omega$ convention. This is
    also the representative mode used in the convergence and
    matrix-pencil analyses, and it supplies the intermediate point
    in the finite-mass $\ell=1$ sequence.

    \item
    \textbf{Konoplya and Zhidenko \cite{Konoplya:2017tvu}.}
    Their Table II gives Frobenius results for the fundamental
    $\ell=2$ massive-scalar mode in terms of the horizon-radius
    variables $\mu r_+$ and $\omega r_+$. Since $r_+=2M$ for
    Schwarzschild, the rows
    $\mu r_+=0.20,0.30,$ and $0.40$ correspond respectively to
    $\mu M=0.10,0.15,$ and $0.20$, while the tabulated frequencies
    are already expressed as $2M\omega$. These entries provide the
    complete $\ell=2$ finite-mass sequence and complement the
    corresponding $\ell=1$ benchmarks.

    \item \textbf{Corr\^ea et al \cite{Correa:2024xki}.}
The quasi-bound-state reference values in
Table~\ref{tab:qbs_frequency_domain_validation} are taken from
Table~II of Ref.~\cite{Correa:2024xki}, which reports
\(\omega_{\rm QBS}/\mu\) at \(\mu r_H=0.5\). For a Schwarzschild
black hole, \(r_H=2M\), so that \(\mu r_H=2\mu M\), and the quoted
mass parameter corresponds to \(\mu M=0.25\). The frequencies are
converted to the convention used here according to
\[
2M\omega_{\rm QBS}
=
(2M\mu)\frac{\omega_{\rm QBS}}{\mu}
=
(\mu r_H)\frac{\omega_{\rm QBS}}{\mu}
=
0.5\,\frac{\omega_{\rm QBS}}{\mu}.
\]
\end{itemize}

\end{document}